\documentclass[useAMS,usenatbib,preprint]{mn2e}

\newcommand{\apj}{ApJ}
\newcommand{\apjs}{ApJ Suppl.}
\newcommand{\apss}{Ap\&SS}
\newcommand{\mnras}{MNRAS}
\newcommand{\araa}{ARA{\&}A}

\newcommand{\apjl}{ApJL}

\newcommand{\nat}{Nature}

\usepackage{amssymb,amsmath}
\usepackage[pdftex]{graphicx}
\usepackage{epstopdf}
\usepackage{times}

\title[Very-Long-Term Variability of XRBs in M87]{Monitoring the Very-Long-Term Variability of X-ray Sources in the Giant Elliptical Galaxy M87}

\author[D. L. Foster et al.]{D. L. Foster,$^{1,2}$\thanks{E-mail: deatrick@saao.ac.za} P. A. Charles,$^{3}$ D. A. Swartz,$^{4}$ R. Misra,$^{5}$ and K. G. Stassun,$^{2,6}$ 
\\
\\
$^{1}$South African Astronomical Observatory, PO Box 9, Observatory, 7935, South Africa\\
$^{2}$Vanderbilt University, Department of Physics \& Astronomy, 1807 Station B, Nashville, TN, 37235, USA\\
$^{3}$University of Southampton, School of Physics \& Astronomy, Southampton, Hampshire, SO17 1BJ, UK\\
$^{4}$Universities Space Research Association, NASA Marshall Space Flight Center, ZP12, Huntsville, AL,  35805, USA\\
$^{5}$Inter University Centre for Astronomy and Astrophysics, Pune University Campus, Pune, India\\
$^{6}$Fisk University, Department of Physics, 1000 17th Avenue North, Nashville, TN,  37208, USA}

\begin{document}

\date{Accepted 2013 March 27. Received 2013 March 26; in original form 2013 February 16}

\pagerange{\pageref{firstpage}--\pageref{lastpage}} \pubyear{2012}

\maketitle

\label{firstpage}

\begin{abstract}
We report on our search for very-long-term variability (weeks to years) in
 X-ray binaries (XRBs) in the giant elliptical galaxy M87.  We have used archival {\it Chandra} imaging observations to characterise the long-term variability of 8 of the brightest members of the XRB population in M87.  The peak brightness of some of the sources exceeded the ultra luminous X-ray source (ULX) threshold luminosity of $\sim 10^{39}$ erg s$^{-1}$, and one source could exhibit dips or eclipses.  We show that for one source, if it has similar modulation amplitude as in SS433, then period recoverability analysis on the current data would detect periodic modulations, but only for a narrow range of periods less than 120 days.  We conclude that a dedicated monitoring campaign, with appropriately defined sampling, is essential if we are to investigate properly the nature of the long-term modulations such as those seen in Galactic sources.

\end{abstract}


\section{Introduction}
{\it Chandra's} sub-arcsecond resolution routinely images tens to hundreds
of X-ray sources in individual observations of galaxies out to Virgo-cluster
distances down to detection limits of $\sim 10^{37}$ erg s$^{-1}$.  In analogy with
the Milky Way, these are mostly X-ray binaries \citep[][]{Fabbiano:2006a, Fabbiano:2006b}.  
Detection limits are of order 10 counts, corresponding to count rates of several $10^{-4}$ c s$^{-1}$ 
for typical exposure times of 10 to 100 ks so that meaningful lightcurves can only be obtained 
for the most luminous sources and only over time intervals of order 1 day at most.  In rare cases,
these lightcurves display periodic dips or other structures that can be used
to constrain properties of the XRBs \citep[e.g.][]{Trudolyubov:2002}. 

Multi-epoch X-ray observations of nearby galaxies allows for the study of the long-term variability of their X-ray source populations.  We know from previous temporal and spectral analysis on the most luminous XRBs in nearby galaxies that they exhibit behaviours similar to what is seen in the Galactic XRBs \citep[][and references therein]{Fabbiano:2006a}.  For example, \cite{Kong:2002} find that among the 204 sources detected in the nearby spiral galaxy M31, 50 per cent are variable on the time-scale of several weeks and 13 are transients.  We expect that ultraluminous X-ray sources \citep[ULXs, defined as extranuclear point-like sources of X-rays with $L_X \gtrsim 10^{39}$ erg s$^{-1}$; e.g.][]{Fabbiano:2006b} exhibit similar very-long-term modulations to those seen in the Galactic XRB population.

Owing to the lack of regular, multi-epoch observations of extragalactic XRBs, not much is known 
about their variability on time-scales of tens to hundreds of days \citep[][]{Fridriksson:2008}.  
Shorter time-scales can also be challenging for a variety of reasons: (a) count
rates are always too low to see very short fluctuations like pulse periods; (b)
a few short, presumably orbital, modulations have been detected in rare cases
in which  the modulations occur 2--3 or more times within a single observation; however, these required exposure times in the 10--100 ks range, yet the typical {\it Chandra} exposure times reported here are less than this; and (c) transient-like
behaviour--in which variations in luminosity are by a factor of 100 or greater--can almost always only
be seen in multi-epoch exposures \citep[typically separated by of order 100 days; e.g.][]{Barnard:2012a}.  Notable exceptions are some of the super-soft sources that vary
rapidly within a given observation as well as between observations \citep[][]{Swartz:2002, Fabbiano:2003, Mukai:2003, Fabbiano:2006a, Orio:2006, Carpano:2007}. Thus, there
are only a handful of examples of category (b) and (c).  However, \cite{Kong:2011} demonstrated that regular pointed X-ray observations can be successfully deployed as monitoring instruments for extragalactic XRBs.  This is particularly true of {\it Chandra X-ray Observatory} because of its superior resolution.

Longer term, non-orbital periodicities have been observed in both LMXBs and HMXBs, are termed {\it superorbital}, and range from tens to hundreds of days \citep*{Wen:2006, Charles:2010}.  The study of these modulations are a probe of the physical processes occurring within both the donor star and the accretion disc, and also a probe of the interaction of the intense X-ray luminosity with the accretion flow.  These superorbital modulations display different properties depending on the type of interacting binary in which they are found \citep[][]{Charles:2010}.  Highly relevant here are systems which display long-term superorbital modulations due to the coupled precession of their accretion discs and relativistic jets, such as the 35-d modulation of \mbox{Her X-1} \citep[][]{Petterson:1977} and the 162-d modulation of SS433 \citep[][]{Margon:1984}; this mechanism may be responsible for the 115-d modulation seen in the ULX \mbox{NGC 5408 X-1} \citep*[][]{Foster:2010}.  Only with a properly designed long-term monitoring program can we distinguish between the competing physical explanations of these superorbital periods.

To date, there have been extensive investigations of low-mass X-ray binaries (LMXBs) in M87 \citep[][]{Jordan:2004b,Jordan:2004}.  Some studies contain observations separated by $\sim$ years, but without regularly repeated visits (typical of monitoring campaigns) extended over a baseline of many years.  These have primarily focused on the pointing-to-pointing variability of the spectra and fluxes of LMXBs, and their association with globular cluster systems within M87. Other important work on these XRBs includes the nature of the apparently years-long transient outbursts discovered in these sources \citep[][]{Irwin:2006}.  Better samplings have been obtained for LMXBs in globular clusters near the central region of M31 \citep*{Barnard:2012b}.

Monitoring of XRBs in the Milky Way has led to numerous major discoveries. For example, the discovery within our own Galaxy of relativistic jets and their superluminal motion in an accreting compact binary \citep[][]{Mirabel:1994} provided a link to the physics of distant quasars that are thought to behave in a similar (yet much more powerful) manner.  The discovery of warping and precessing of the accretion discs was revealed through monitoring the long-term periodic variations in systems like SS433, \mbox{Her X-1}, and \mbox{LMC X-3} \citep*[][]{Margon:1980, Cowley:1991, Gerend:1976}.  And, the discovery of sub-millisecond quasi-periodic oscillations (QPOs) in \mbox{Sco X-1} \citep[][]{Klis:1996} demonstrated a probe of the dynamics of the inner accretion flow in a XRB.  As was shown in \cite{Margon:1980}, only with regular, sustained monitoring is it possible to measure longer-term (quasi-) periodicities that range from weeks to hundreds of days, with greater precision for the longer variations such as that achieved for SS433 coming only after years (up to decades) of follow-up observations \citep{Eikenberry:2001}.

{\it Chandra} has now studied M87 many times over $\sim$ 10 years; this has the potential for showing long-term modulations or transient behaviour in several XRBs in M87 \citep[NGC 4486, $D = 17$~Mpc;][]{Tully:2008}.  We have selected a handful of the brightest sources in this galaxy and searched archival data to study their very long-term variability.

\section{Methods}

\subsection{Observations and data reduction}
The {\sc CIAO} data reduction threads were used to process the data, generating level-2 events lists for each 
observation.  We used version 4.4.1 of {\sc CIAO} and version 4.5.1 of {\sc CALDB} to reprocess the data.  All the data presented here were recorded with {\it Chandra's} Advanced CCD Imaging Spectrometer (ACIS) S3 chip.

The data span the period 2000 July 29 to 2010 May 14.  With a 
handful of exceptions, the 85 independent observations included in this analysis were typically 5 ks in 
duration.  Chandra detects photons at a rate of about $10^{-4}$ c s$^{-1}$ for a source of about $10^{-15}$ erg cm
$^{-2}$ s$^{-1}$ or only about 15 counts from a borderline ULX ($\sim 10^{39}$ erg s$^{-1}$) in these 5 ks observations.  
Also, many of the observations were performed in subarray mode (mostly due to the brightness of the core 
and jet of M87) and with varying roll angles.  This led to somewhat sporadic coverage for a few sources.  

To test for the presence of periods of high background or strong flares, we excluded the regions of the chip containing the core and jet of M87, as well as regions containing point sources detected by the {\sc CIAO} tool \verb"wavdetect", then examined the lightcurve of the remaining events of the observation (binning of 200s) using the tool 
\verb"lc_sigma_clip", rejecting any events that are beyond 5$\sigma$ from the mean count rates.  No significant flares or periods of high background are detected.  This means that the total exposure accumulated during the good time intervals (GTIs) as recomputed by \verb"lc_sigma_clip" are nearly identical to the value of the keyword \verb"EXPOSURE" in each observation's FITS file, and this time was used to compute the mean count rate.

\subsection{Source selection}
Eight sources were visually selected for analysis based on their apparent relative brightness in the X-ray images, their higher sampling given the roll angles of the observations,  and their being distinct from either the nucleus of the galaxy or the jet (including knots or other clumps of diffuse emission). All of these sources appear in the catalogue of sources analysed in \cite{Jordan:2004}.  The X-ray sources are listed in Table \ref{summary}, and are hereafter individually referred to by the labels indicated therein (X1, X2, et cetera).

\subsection{Extraction of source count rates}
Source regions were defined as 1.5-arcsec circles centred on the identified point sources.  The background regions were defined for the sources as annuli centred on the sources with inner and outer radii of 1.5 and 2.5 arcsec, respectively, to account for the diffuse emission that surrounds some of the sources while minimizing the effects of overlapping source and background regions.  A spatial extraction of the total number of counts in the source and background regions, in the 0.5--8.0 keV range, was performed using the {\sc CIAO} tool \verb"dmextract", with the error in the counts estimated using Gehrels' method \citep[][]{Gehrels:1986}.  The larger of the asymmetric Poisson errors is used for all calculations, so as to be conservative.  Each point on the lightcurves represents the mean count rate (in c s$^{-1}$) of an individual observation.  Data points on the lightcurves that are consistent with zero counts per second to within twice their 1$\sigma$ Poisson errors are denoted by arrow markers on all lightcurves, and we treat these data as upper limits.  An example is shown in Figure \ref{lightcurves}, which is the long-term lightcurve for the source X7.

The largest distance of any of these sources from the focal point (i.e. the narrowest point-spread function, or PSF) of the ACIS S3 chip is $\approx 3$ arcmin.  We estimate the 90 per cent encircled energy radius of a 1.5-keV event to be 1.0 arcsec on-axis and 2.0 arcsec at 3 arcmin off-axis, and that for a 6.4-keV event to be 2.0 arcsec on-axis and 2.6 arcsec at 3 arcmin off-axis.  We checked the most off-axis sources for any differences in period detection arising from the factor of 2 increase in the radii of the extraction apertures and background annuli, finding none.  In the event that a particular source was not on the active subarray, there was no data point recorded for such a source.  We confirmed each off-subarray instance visually with \mbox{{\sc SAOImage DS9}} after inspecting all extractions returning exactly zero counts for both source and background regions.

\section{Analysis \& Results}
\label{analysis}

\subsection{Peak X-ray Flux and X-ray Luminosity}
\label{flux}
Table \ref{summary} shows the observations having the largest X-ray flux and luminosity for each source. $F_{\rmn{max}}$ (in units of $10^{-14}$ erg cm$^{-2}$ s$^{-1}$) and $L_{\rmn{max}}$ (in units of $10^{39}$ erg s$^{-1}$) are the maximum values of the net model-independent (i.e., directly summed) X-ray flux and luminosity, respectively, over the entire baseline of observations.  The sum of the flux within each source and background region was computed using the {\sc CIAO} tools \verb"eff2evt" and \verb"dmstat".  The source regions are defined exactly as in the lightcurve extraction step and  background fluxes were estimated for a region of the same radius near the source using \mbox{{\sc SAOImage DS9}}.  The events were filtered on the 0.5--8 keV band. The fluxes are estimates based on the most likely energy of each arriving photon, and do not take into account a spectral model and the redistribution matrix.  The luminosities are scaled to the assumed distance to M87 of 17 Mpc and uncorrected for absorption.  All of our analysis below uses only the relative fluxes to search for variability and periodicity.  The OBSID of the data sets corresponding to the peak values $F_{\rmn{max}}$ and $L_{\rmn{max}}$ for each point source is given as a reference.

\subsection{Statistical analysis}
Having invoked the Gehrels error estimation method in the counts extraction step, the Poisson errors in the case of low counts are properly taken into account. To be conservative, the larger of the asymmetric Poisson errors is used.

Each lightcurve was fitted with a horizontal line representing the median count rate for the individual sources to determine how well they agree with being constant.  The results are shown in Table \ref{summary}.  In the goodness-of-fit analysis for $\chi^2$, the null hypothesis is rejected for each source except for X3, indicating that the sources are very unlikely to be non-variable.

\begin{figure}
\includegraphics[width=90mm]{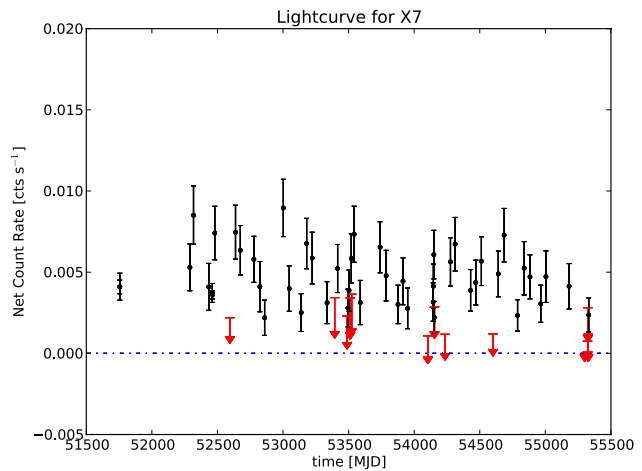}
\caption{Long-term lightcurve for X7. Events were extracted in the 0.5--8.0 keV band.  Each point on the lightcurves represents the mean count rate (in c s$^{-1}$) of an individual observation.  The level of zero counts per second is represented by the dashed line.  Arrow markers represent data that are consistent with zero counts per second, and the arrow length is the measurement error associated with the particular data point.  In our analysis, we treat these points as upper limits.  Appendix A contains the information for all the sources.}
\label{lightcurves}
\end{figure}

\subsection{Period search}
\label{periodsearch}
\begin{figure}
\includegraphics[width=90mm]{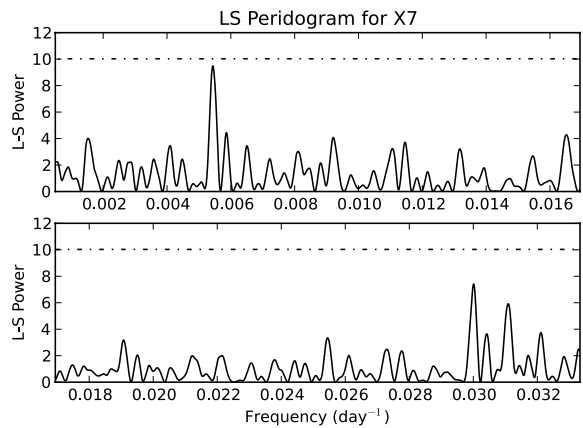}
\caption{Lomb-Scargle periodogram for the data plotted in Figure \ref{lightcurves}. The frequencies correspond to periods ranging from 30 d to 5 yr.  The highest peak corresponds to a period of $\sim$ 180 d. The dashed line represents the 3$\sigma$ Monte Carlo confidence level.  Appendix B contains the information for all the sources.}
\label{lsp}
\end{figure}

\begin{figure}
\includegraphics[width=90mm]{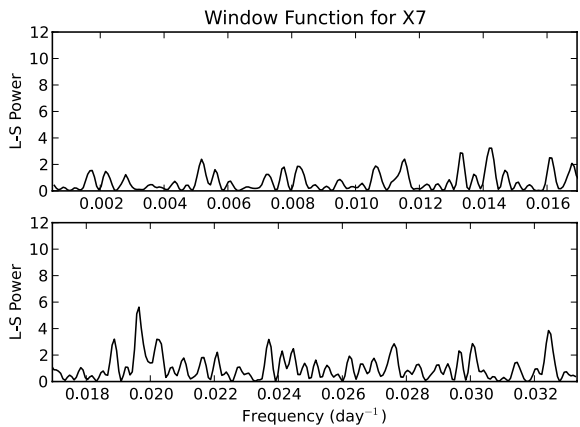}
\caption{Window function for the data plotted in Figure \ref{lightcurves}. The frequencies correspond to periods ranging from 30 d to 5 yr.  Appendix C contains the information for all the sources.}
\label{windowN96}
\end{figure}

Lomb-Scargle periodogram (LSP) analysis \citep[][]{Lomb:1976, Scargle:1982, Press:1989} was applied to all extracted lightcurves and the resulting power spectra were searched for periods ranging from 30 days to 5 years.  Figure \ref{lsp} is the LSP of the data in Figure \ref{lightcurves}.  In order to identify spurious period detections due to sampling, each power spectrum was compared to its corresponding window function power spectrum.  Figure \ref{windowN96} is the window function of the data in Figure \ref{lightcurves} corresponding to the power spectrum in Figure \ref{lsp}.

To determine the white noise levels, Monte Carlo simulations of 10,000 randomly generated data sets for each source were made using the time values, and the mean and standard deviation of the count rate values.  LSPs were computed for the random data sets, and the resulting maxima in the power spectra are used to construct a cumulative probability distribution function, which describes the false-alarm probability (FAP) associated with each value of the power.  Hence, a 3$\sigma$ confidence level is the power associated with FAP of 0.27 per cent.  For our lightcurves, the power typically associated with FAP of 0.27 per cent is $\approx$ 6--10.  For the source X6, the number of data points is very small, and the computed 3$\sigma$ confidence level is suspiciously low and perhaps not the ``true'' value.

\subsection{Period recoverability analysis}
\begin{figure*}
\includegraphics[width=130mm, angle=90]{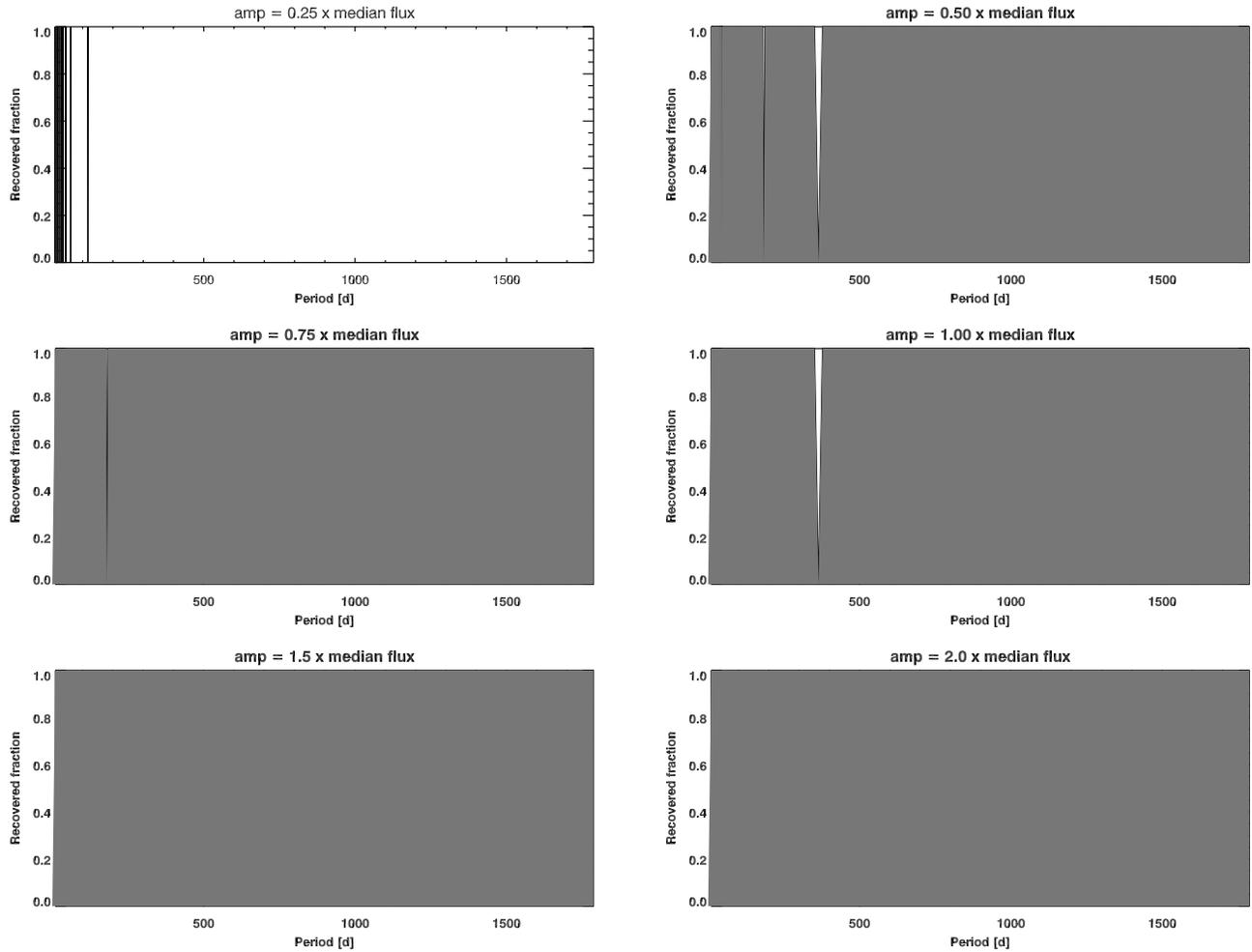}
\caption{Period detection sensitivity for X7. We use the times and counts errors of the data points of the extracted lightcurve to generate 10,000 simulated lightcurves per period bin with randomized phases. LSP analysis was performed to test whether the input periods are recovered.  The recovery fraction of each input period is determined by evaluating the LSPs of all the simulated lightcurves in each period bin. Each panel represents a different fixed input modulation amplitude.  The recovery fraction is indicated by grey shading.  Agreement to within 10 per cent of the input period is deemed a successful recovery.    Appendix D contains the information for all the sources.}
\label{recover}
\end{figure*}

Following an identical period search algorithm in section \ref{periodsearch}, we used our observed lightcurves to perform Monte Carlo simulations to calculate the recoverability of an extended range of periods.  The recoverability chart for the data plotted in Figure \ref{lightcurves} is shown in Figure \ref{recover}.  We conducted period recoverability analysis for amplitudes ranging from 25--200 per cent for the source X7.  Generally, the recoverability charts show the expected result that the sampling of the available data is adequate for recovery of longer-term periods for higher amplitude modulations and for higher count rates.  Interestingly, the charts also show that for amplitudes of $\sim$ 20 per cent, which is similar to that found in SS433, long-term periods would have low recoverability with the current sampling, and hence we would not have expected to see them had they been present.  However, a more extreme amplitude variation of  $\gtrsim$ 50 per cent could have enabled recovery of a wide range of periods in the existing data for the source X7 had they been present.

\section{Discussion}
None of the sources show strong evidence for periodic variations, although the source X7 does show marginally significant periodic variability.  The remaining sources may show variability of a more general nature (see Table \ref{summary}).  Henceforth, we will use the case of X7 (see Figure \ref{lightcurves}) as an exemplar in our discussion, although the analysis discussed in Section \ref{analysis} was applied to all sources.

\begin{figure*}
\includegraphics[width=84mm]{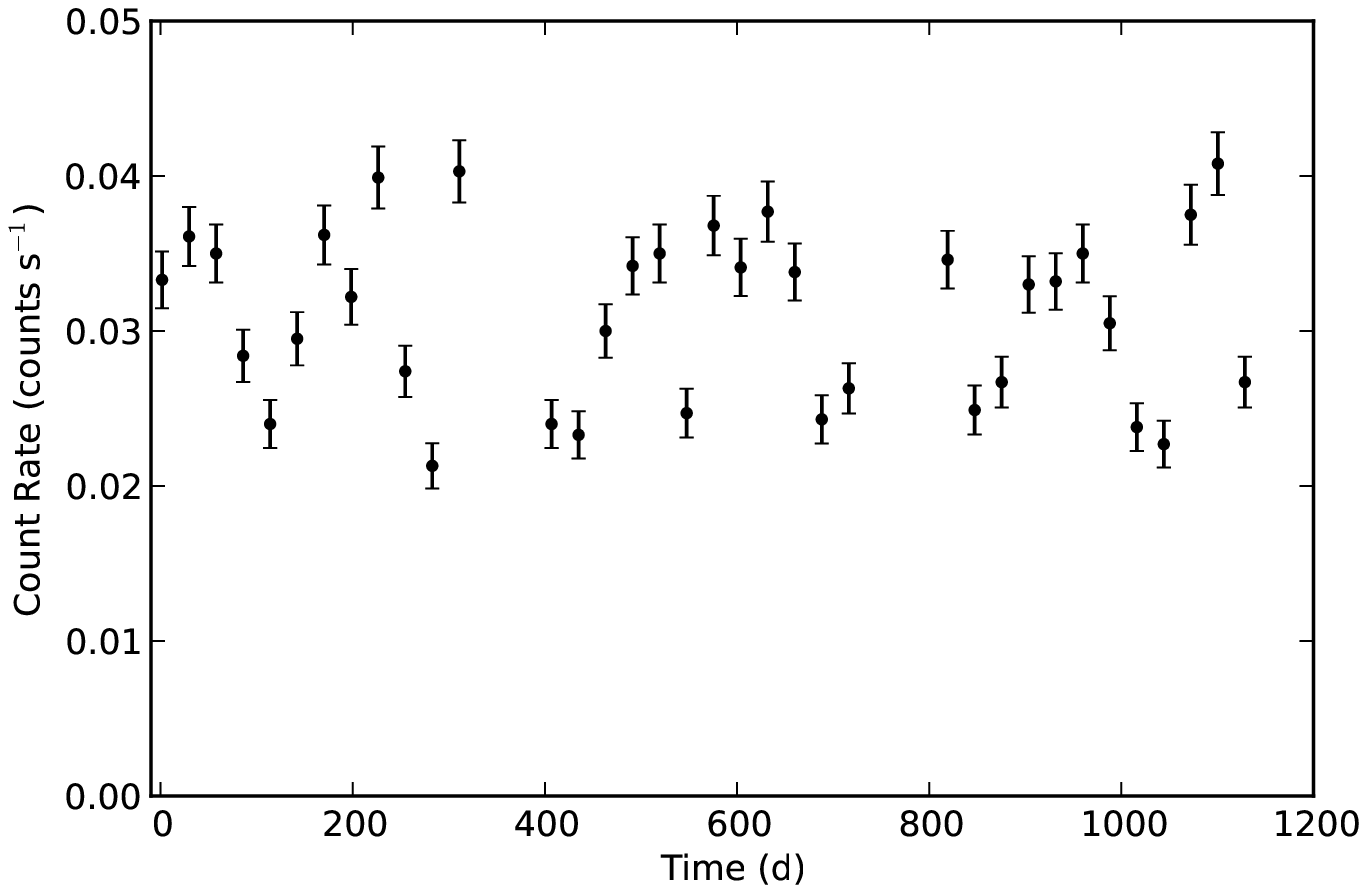}
\includegraphics[width=84mm]{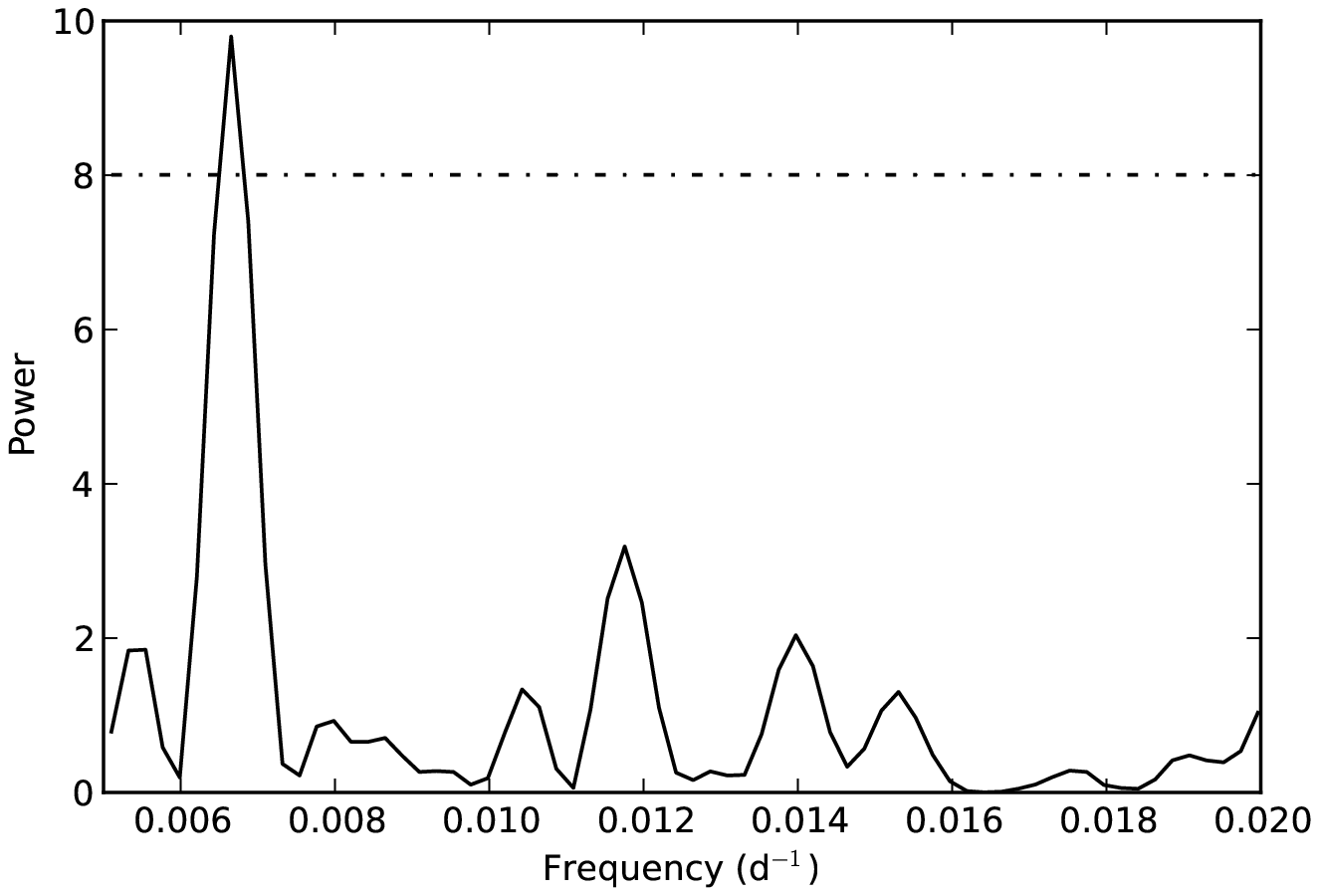}
\caption{{\it Left panel}:  Simulated {\it Chandra} ACIS light-curve of a ULX with a mean count rate of 0.03 c s$^{-1}$, which exhibits a 150-d sinusoidal modulation of 20 per cent.  Twelve observations (one every 4 weeks) are simulated during each year for 3 years.  {\it Right panel}:  Lomb-Scargle periodogram of the data plotted to the left, together with the 3$\sigma$ Monte Carlo confidence level (dashed line).}
\label{ulxsim}
\end{figure*}

\begin{figure*}
\includegraphics[width=84mm]{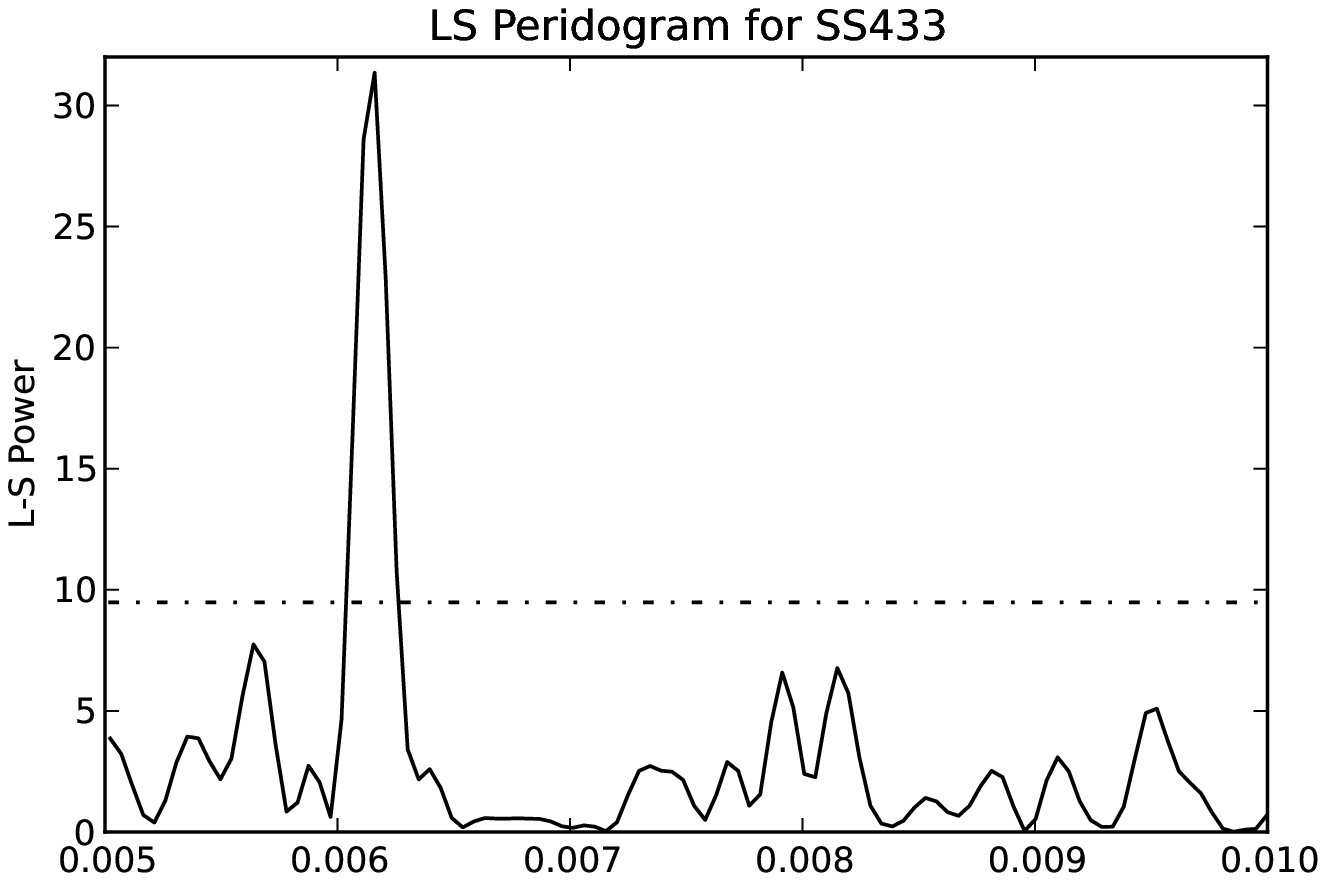}
\includegraphics[width=84mm]{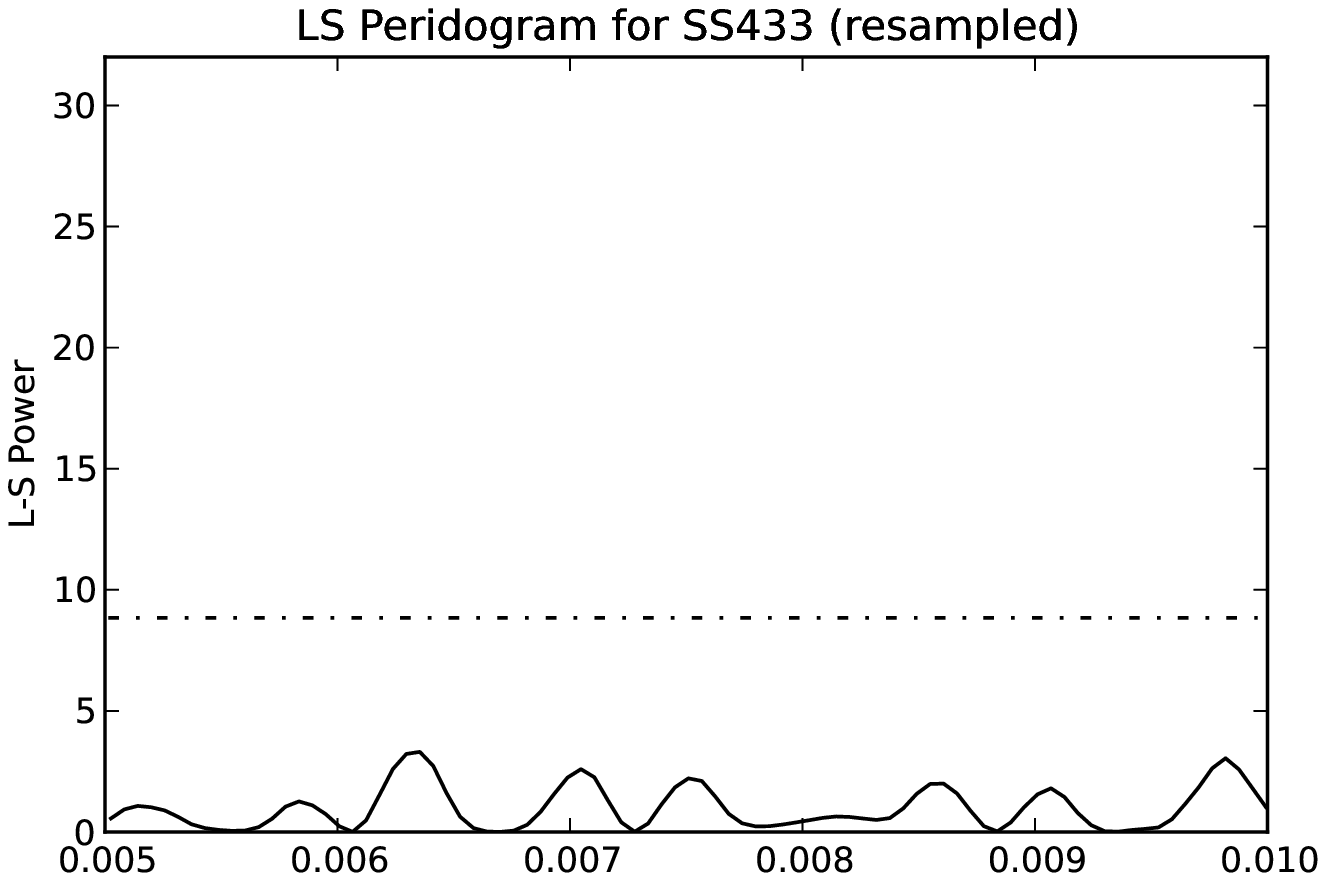}
\caption{{\it Left panel}: Lomb-Scargle periodogram of the full 15-year RXTE/ASM 1-day averages lightcurve of SS433, which exhibits a $\sim$160-d sinusoidal modulation.  {\it Right panel}:  Lomb-Scargle periodogram of the same data resampled to match the sampling of X1, one of the best sampled sources in our analysis.  The dashed lines represent the 3$\sigma$ Monte Carlo confidence limits, illustrating that the random sampling has missed a known long-term period.}
\label{ss433resample}
\end{figure*}

\begin{table*}

\caption{Summary of M87 X-ray source properties}
\label{summary}

\begin{tabular}{ccccc|||ccc}
\multicolumn{2}{c}{} & \multicolumn{3}{c}{Observation-specific properties} & \multicolumn{3}{c}{General properties} \\
\hline
No. & Name & $F_{\rmn{max}}$ & $L_{\rmn{max}}$ [D=17Mpc] & OBSID & Median Rate & $\chi^2/dof$ & $p$\\
   &   (CXOU) & (10$^{-14}$ erg cm$^{-2}$ s$^{-1}$) & (10$^{39}$ erg s$^{-1}$) & & ($10^{-3}$ c s$^{-1}$) & & \\
\hline

X1 & J123047.7+122334 & 12.56 & 4.34 & 8576 & 3.42 & 178/79 & $< 10^{-4}$\\ 

X2 & J123049.2+122334 & 6.81 & 2.36 & 8513 & 3.69 & 193/79& $< 10^{-4}$\\ 

X3 & J123049.6+122333 & 1.73 & 0.60 & 8516 & 1.18 & 79/79 & 0.54\\ 

X4 & J123047.1+122415 & 8.98 & 3.10 & 8513 & 3.51 & 1199/32 & $< 10^{-4}$\\ 

X5 & J123053.2+122356 & 2.68 & 0.93 & 11513 & 1.01 & 115/37 & $< 10^{-4}$\\ 

X6 & J123044.7+122434 & 3.15 & 1.09 & 2707 & 0.56 & 297/7 & $< 10^{-4}$\\ 

X7 & J123050.8+122502 & 6.54 & 2.26 & 4918 & 4.09 & 577/59 & $< 10^{-4}$\\ 

X8 & J123049.1+122604 & 4.40 & 1.52 & 2707 & 3.64 & 557/13 & $< 10^{-4}$\\ 

\hline
\end{tabular}

\medskip
Maximum flux and luminosity (at the distance to M87) for each source, calculated as described in section \ref{flux} and uncorrected for absorption.  The observation identification numbers (OBSID) corresponding to the peak values $F_{\rmn{max}}$ (in 10$^{-14}$ erg cm$^{-2}$ s$^{-1}$) and $L_{\rmn{max}}$ (in 10$^{39}$ erg s$^{-1}$) for each source are shown for reference.  Each source's data are fitted to a constant (the median count rate), and the resulting $\chi^2$ statistic is tabulated, along with the degrees of freedom $(dof)$. $p$ is the null hypothesis probability.

\end{table*}

\subsection{The need for a long-term \textbfit{Chandra} campaign}
The nature of ULXs in nearby galaxies continues as a subject of intense speculation and debate.  Once considered to be the long-sought intermediate-mass BHs, current models invoke slightly heavier than stellar-mass BHs (20--50M$_\odot$) to account for their observed spectra and luminosities \citep{Gladstone:2009}.  Their controversial status has been maintained with contemporary mass estimates for the ULX \mbox{NGC 5408 X-1} ranging from 20 to 5000M$_\odot$ \citep{Strohmayer:2009}.  Its 115-d modulation is either orbital, or superorbital and related to a precessing jet, as in the hyperaccreting SS433.  If the latter is the case then ULXs might be expected to display superorbital variations in the 50--200 day range, and these should be visible in X-ray lightcurves.  Such variations require multi-year long term systematic studies and could only be carried out by {\it Chandra} due to its superior spatial resolution and its more flexible scheduling capabilities compared with other X-ray telescopes.  Rather than having random sampling of the XRBs, if we design a sampling plan such as is proposed here (section \ref{propsim}) then we are able to recover these long-term periods.

\subsection[]{Simulation of a long-term variability}
\label{propsim}
To demonstrate the potential of a long-term X-ray campaign, we have used the observed X-ray modulations of NGC 5408 X-1 and SS~433 (described above as indicative of the periodicities we wish to search for) as input to simulate the outcomes of such a campaign.  The range that we wish to be sensitive to (40--200 days) infers a data sampling rate of 12 observations per year (i.e. one every 3--4 weeks), which is then repeated over the subsequent two years.  We found that the observations within each year's observing window can be pseudo-random, in that they should not be closer together than 14 days.

To test the viability of such a program, we have simulated the expected lightcurves for ULXs having a mean count rate of 0.03 counts per second. A time series is generated with an aperiodic, white noise component with root mean square variability $rms_{N}$ to which we add a coherent sinusoidal signal with amplitude $A$ and period $P$.  The average of the series is taken to be the actual detected count rate of the ULX being simulated. The time series is binned over the typical observation duration of $\Delta T = 10^4$ s
and a Poisson realisation is undertaken to obtain a simulated light-curve of counts per time bin. The simulated light-curve is then sampled on the 36 observations spread over the 3 years, and these data points are input to periodogram analysis, in order to see whether the input periodicity $P$ can be recovered.

We have constructed simulations with periodic modulations of 50, 100 and 150 days, with amplitudes of 20 per cent (typical of \mbox{M82 X-1}, \mbox{NGC 5408 X-1}, and SS433), and for a variety of samplings.  All recover the input periodicity, but with varying degrees of significance.  Best results are obtained with 12 observations per year (spacings of 2--4 weeks), for all the periodicities, and a typical example, including both the lightcurve and the power spectrum, is shown in Figure \ref{ulxsim}.  Compare this with the power spectrum of the 15-year {\it RXTE}/ASM 1-day averages data \citep[][]{Wen:2006} for SS433 in Figure \ref{ss433resample}, which shows that if we have the random sampling of the M87 data (right panel), the well-established 162-d period \citep[e.g.][]{Eikenberry:2001} is not recovered.

All previous attempts to study long-term variability with {\it Chandra} and/or {\it XMM-Newton} have been based on relatively few archival observations randomly distributed in time.  They serve only to indicate why such a study needs to be undertaken in a well-defined, systematic manner over at least a 3-year baseline, as simulated here.

\section{Conclusions \& Future Work}
We identified 8 XRBs in M87 to search for very-long-term periods ($P \sim$ weeks to years).  Using Lomb-Scargle periodogram analysis, we found no statistically significant periods.  We also found no visible evidence of transient outburst events.  However, using period recoverability analysis we found that for amplitude modulations of $\sim$ 20 per cent (similar to what is observed for SS433, NGC5408, and M82), period recovery with the current data is very low for a wide range of periods.  Only if the amplitude of the modulation is $\gtrsim$~50 per cent could a wide range of periods be found.

We continue to monitor the {\it Chandra} archive for sources observed over many years.  Perhaps with better sampling of these sources with facilities such as {\it Chandra}, we can begin to understand their longer-term variations and its consequences for the dynamics of XRBs.

\section*{Acknowledgments}
We thank the referee for providing thoughtful suggestions which have improved this article.  This research has made use of \mbox{{\sc SAOImage DS9}}, developed by Smithsonian Astrophysical Observatory.  The {\sc STARLINK} package \verb"PERIOD" was used to make all periodograms and window functions.  Automation of the analysis process was achieved with scripts written in {\sc PYTHON}, and \verb"matplotlib" was used to generate the figures included with this article.

We thank Dr Alan Levine for providing the {\sc FORTRAN} program used to obtain the white noise estimates for the significance tests.  Furthermore, we thank Dr Mark Finger, Dr Allyn Tennant, and Dr Dan Harris for many helpful comments and suggestions.

DLF acknowledges support from NASA in the form of a Harriett G. Jenkins Fellowship; the Vanderbilt-Cape Town Partnership Program; the Astrophysics, Cosmology \& Gravity Centre (ACGC) at the University of Cape Town; and the International Academic Programmes Office (IAPO) at the University of Cape Town.\\
\indent {\it Facilities: CXO} (ACIS), {\it RXTE} (ASM).

\footnotesize{

}

\appendix

\section{Lightcurves}

\begin{figure*}
\includegraphics[width=84mm]{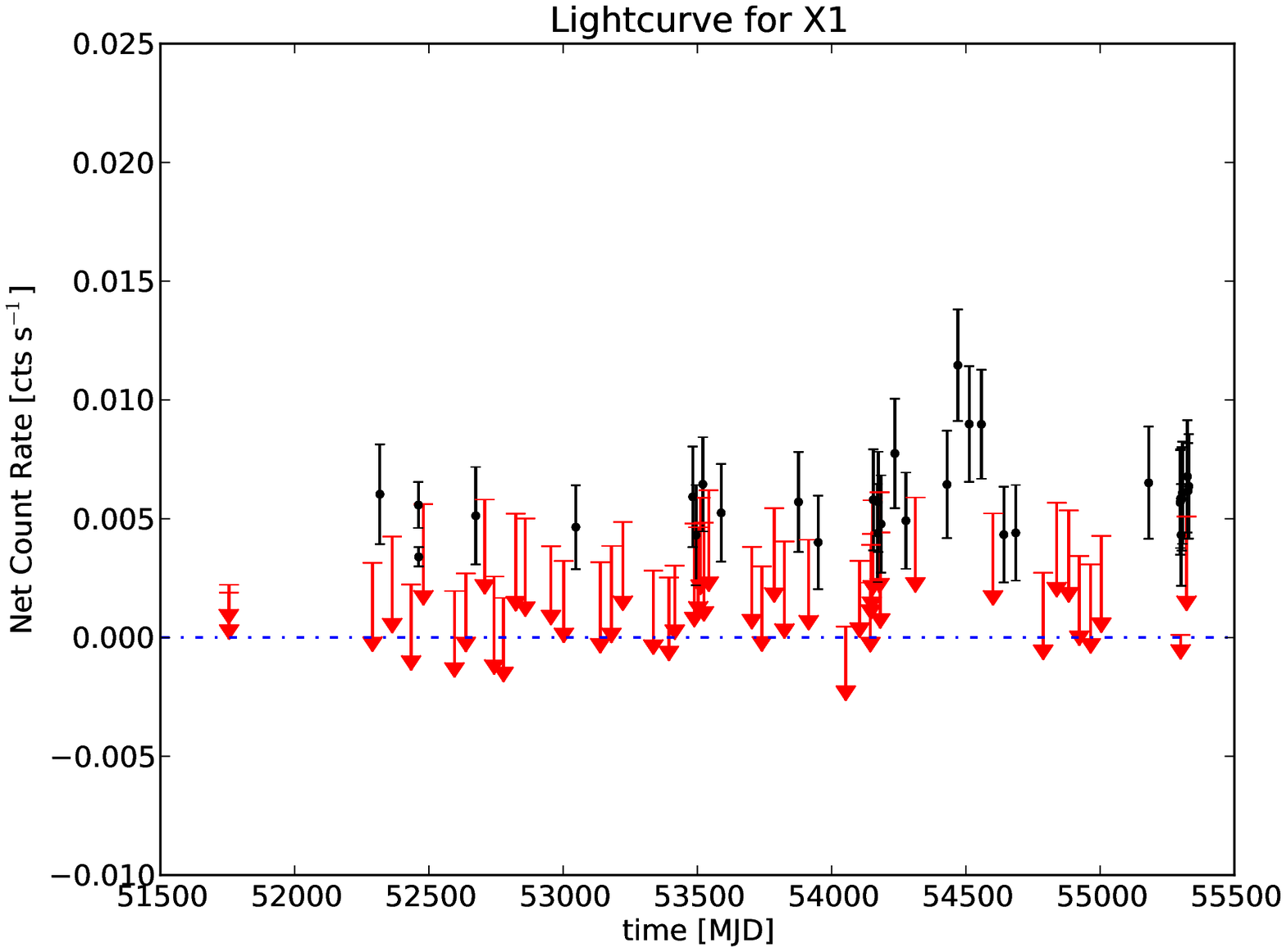}
\includegraphics[width=84mm]{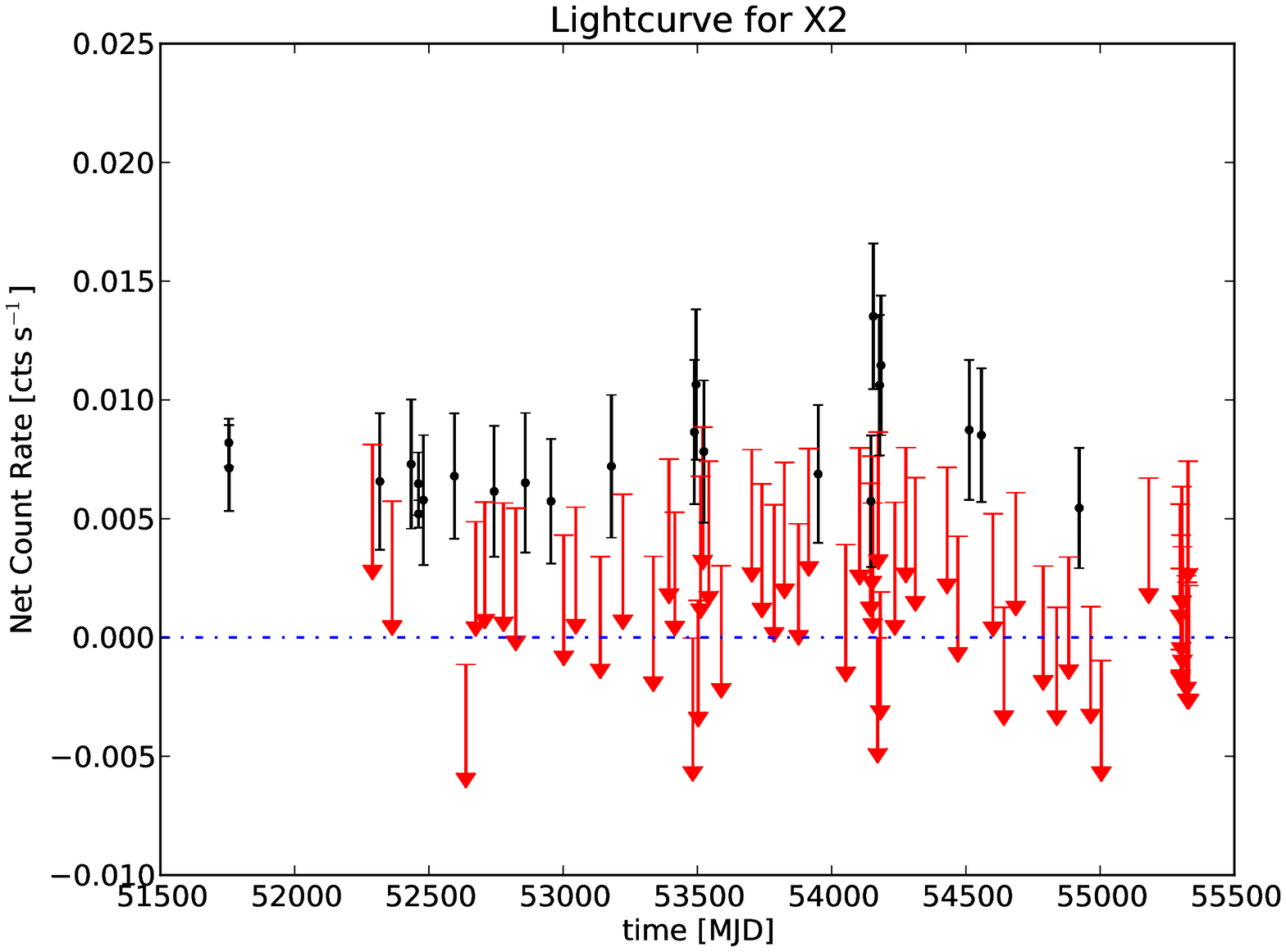}
\includegraphics[width=84mm]{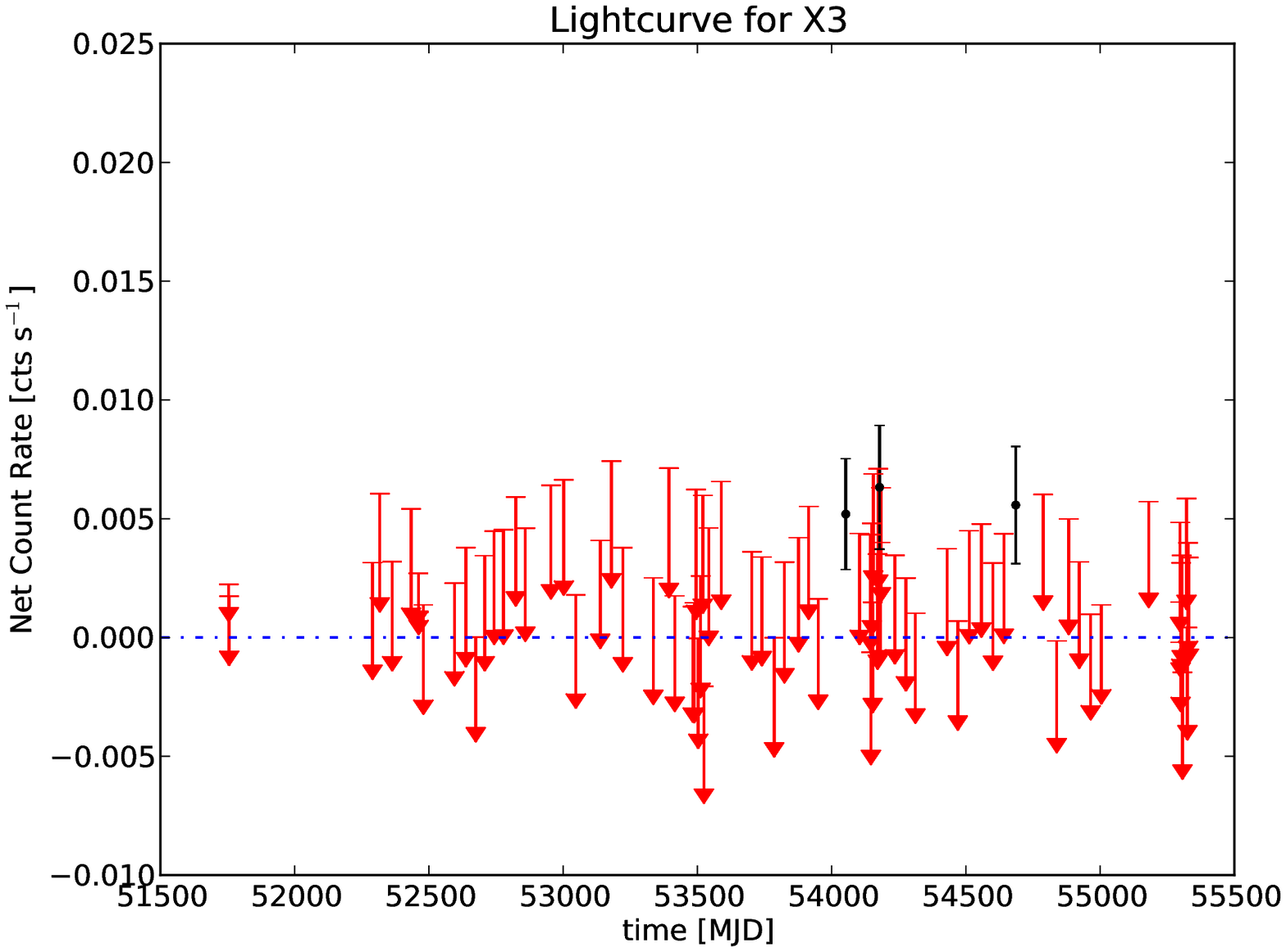}
\includegraphics[width=84mm]{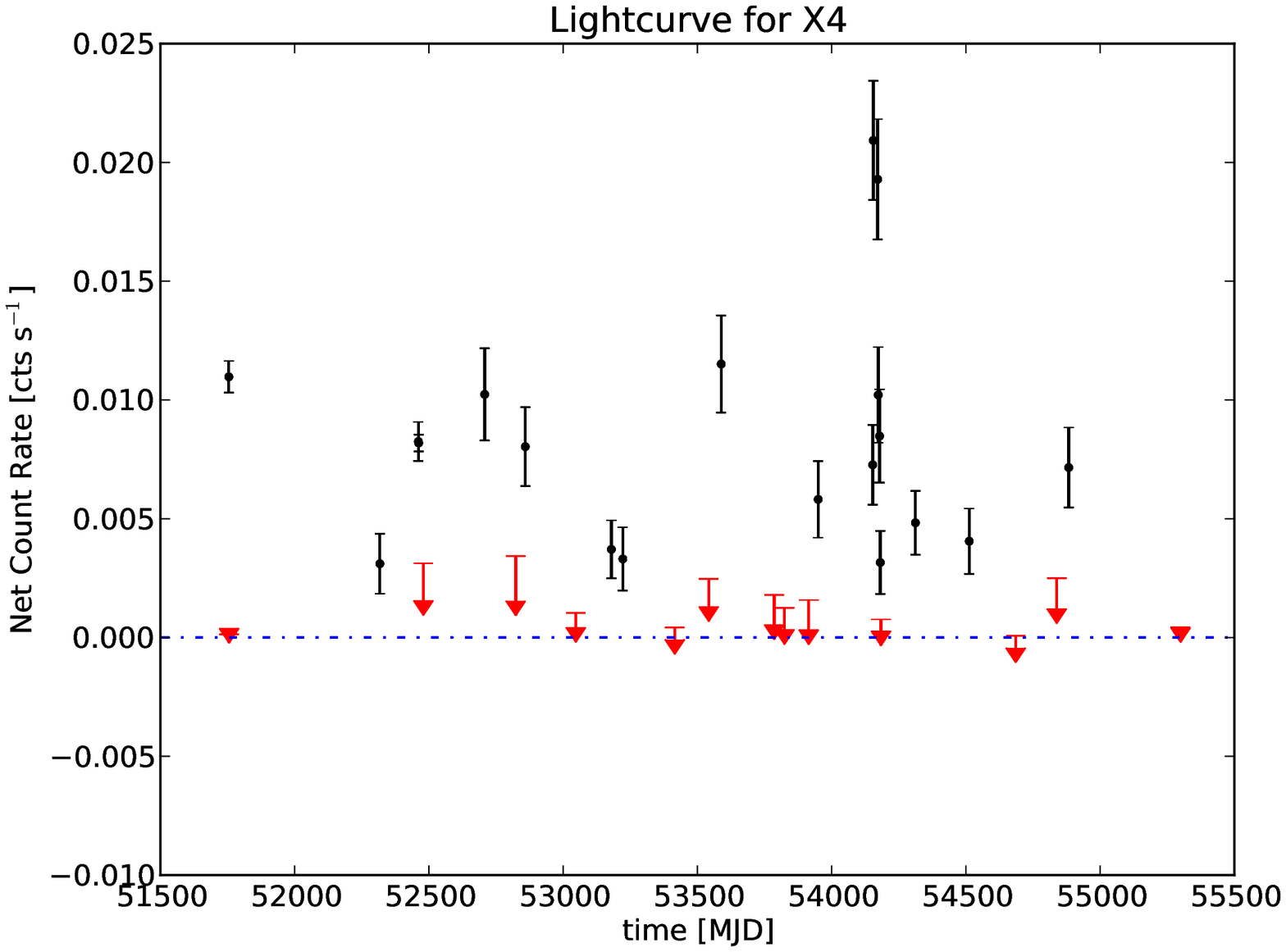}
\caption{Long-term lightcurve for sources X1, X2, X3, and X4. Events were extracted in the 0.5 - 8.0 keV band.  Each point on the lightcurves represents the mean count rate (in c/s) of an individual observation.  The level of zero counts per second is represented by the dashed line. Arrow markers represent data that are consistent with zero counts per second, and the arrow length is the measurement error associated with the particular data point.  In our analysis, we treat these points as upper limits.}
\label{othercurves1}
\end{figure*}

\begin{figure*}
\includegraphics[width=84mm]{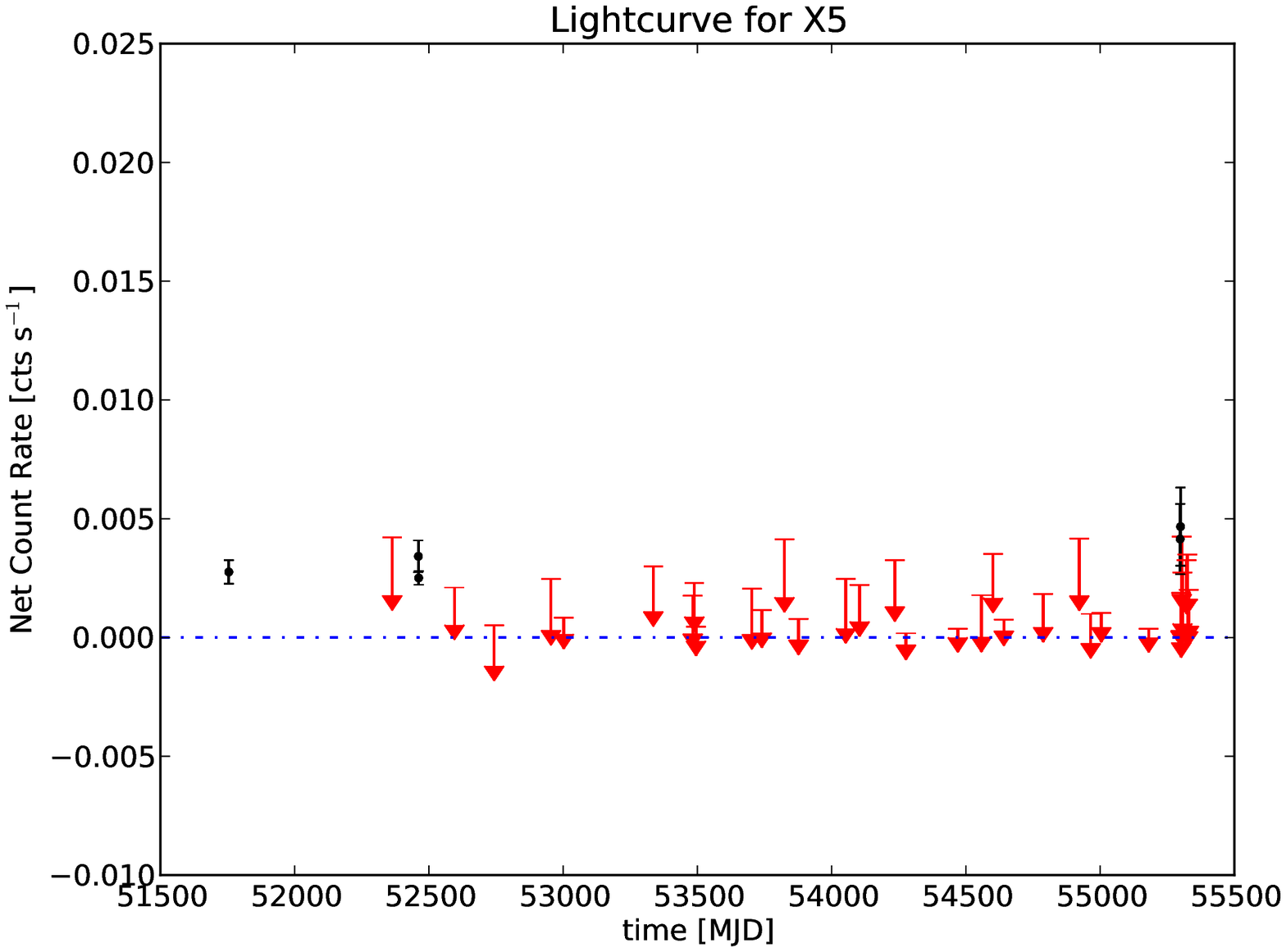}
\includegraphics[width=84mm]{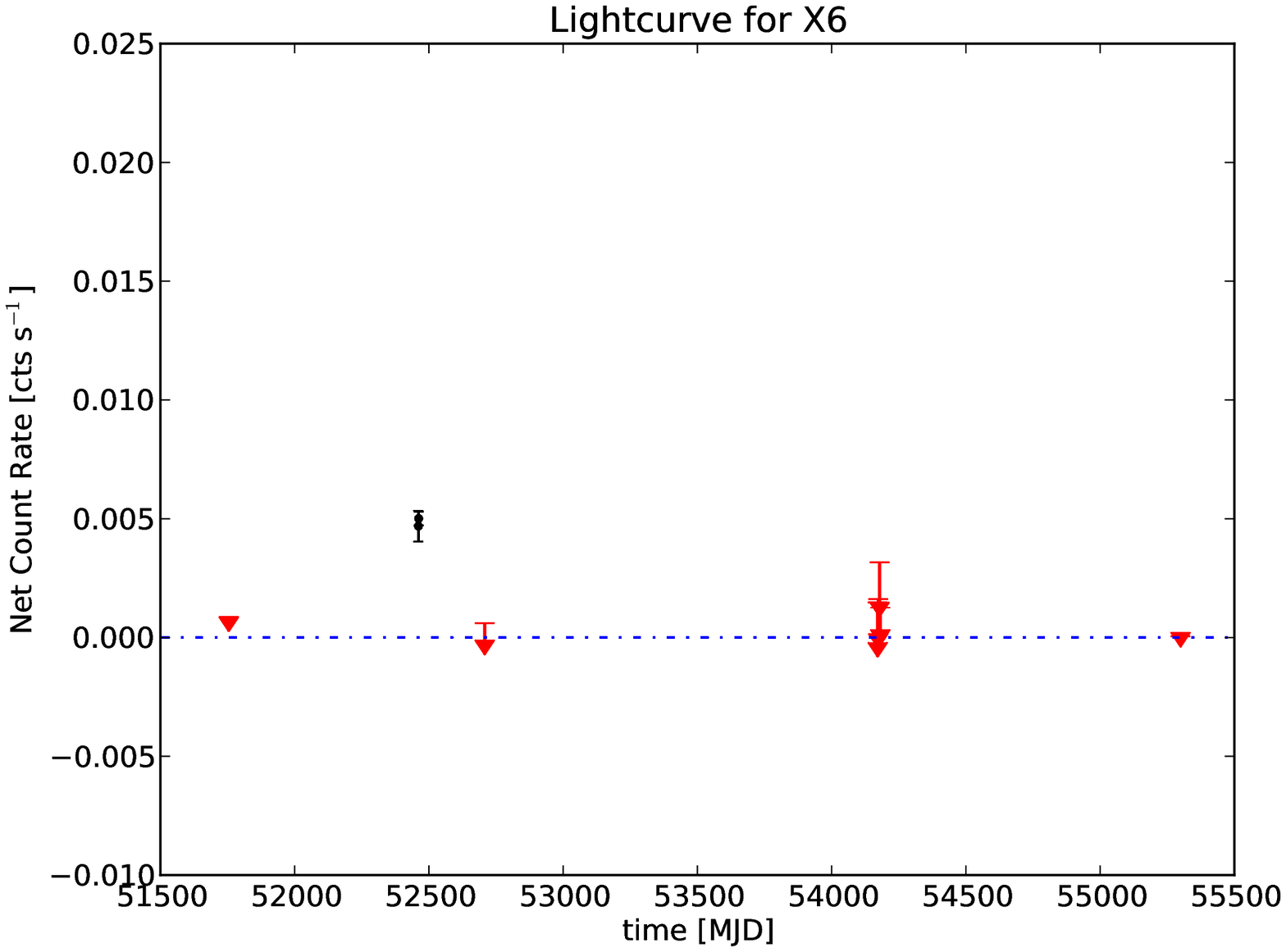}
\includegraphics[width=84mm]{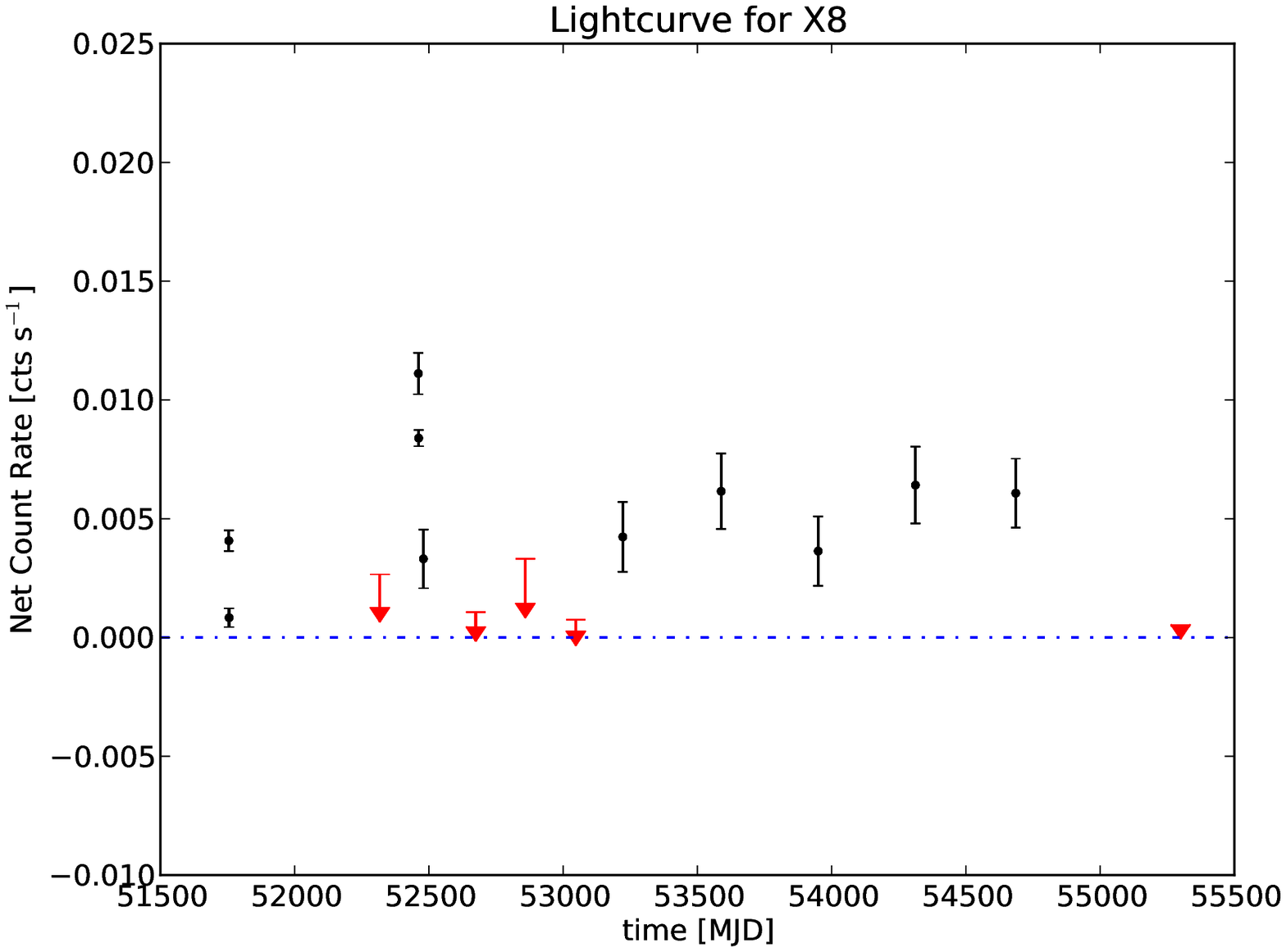}
\caption{Long-term lightcurve for sources X5, X6, and X8. Events were extracted in the 0.5 - 8.0 keV band.  Each point on the lightcurves represents the mean count rate (in c/s) of an individual observation.  The level of zero counts per second is represented by the dashed line. Arrow markers represent data that are consistent with zero counts per second, and the arrow length is the measurement error associated with the particular data point.  In our analysis, we treat these points as upper limits.}
\label{othercurves2}
\end{figure*}

\section{Lomb-Scargle Periodograms}

\begin{figure*}
\includegraphics[width=80mm]{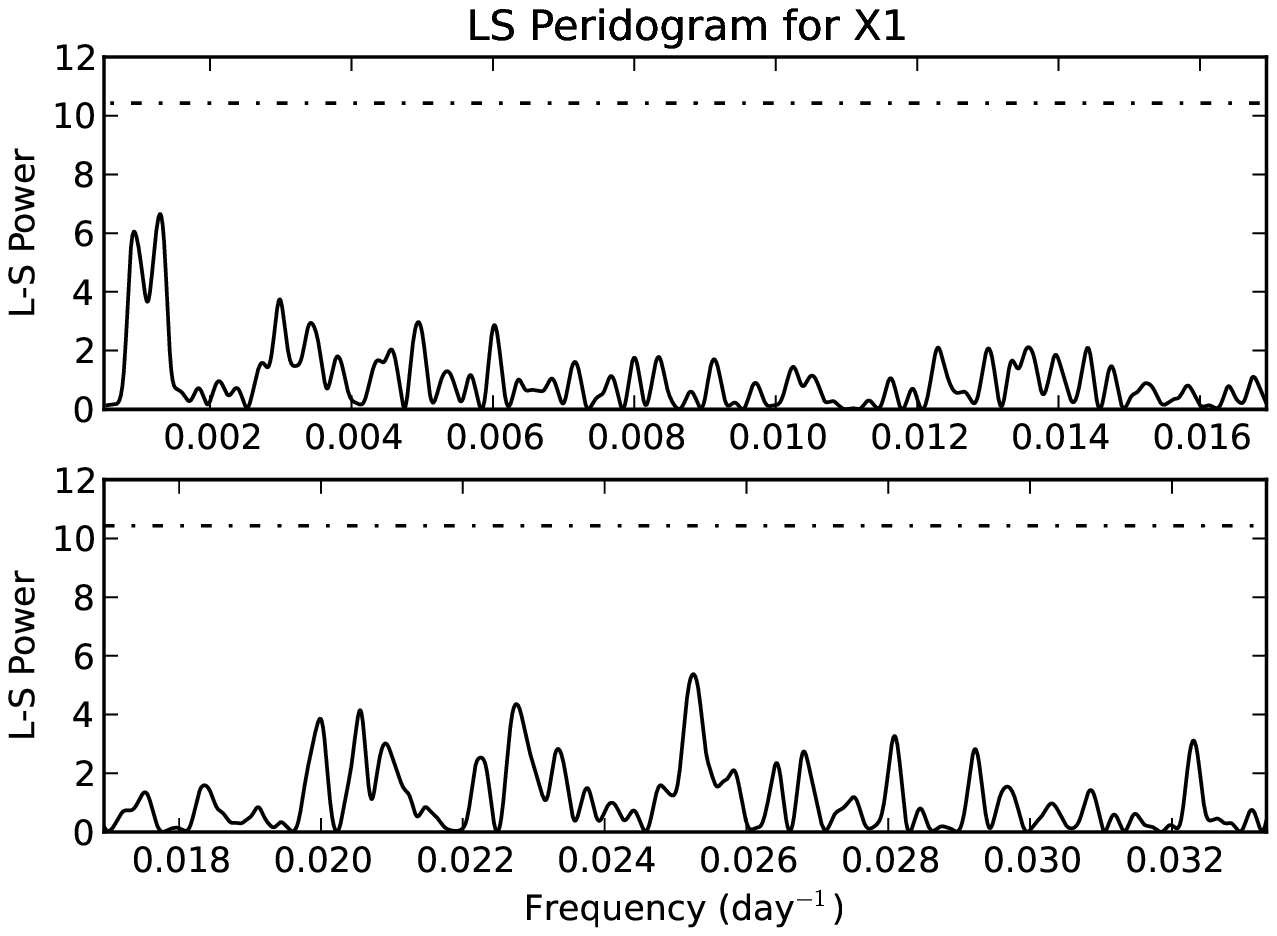}
\includegraphics[width=80mm]{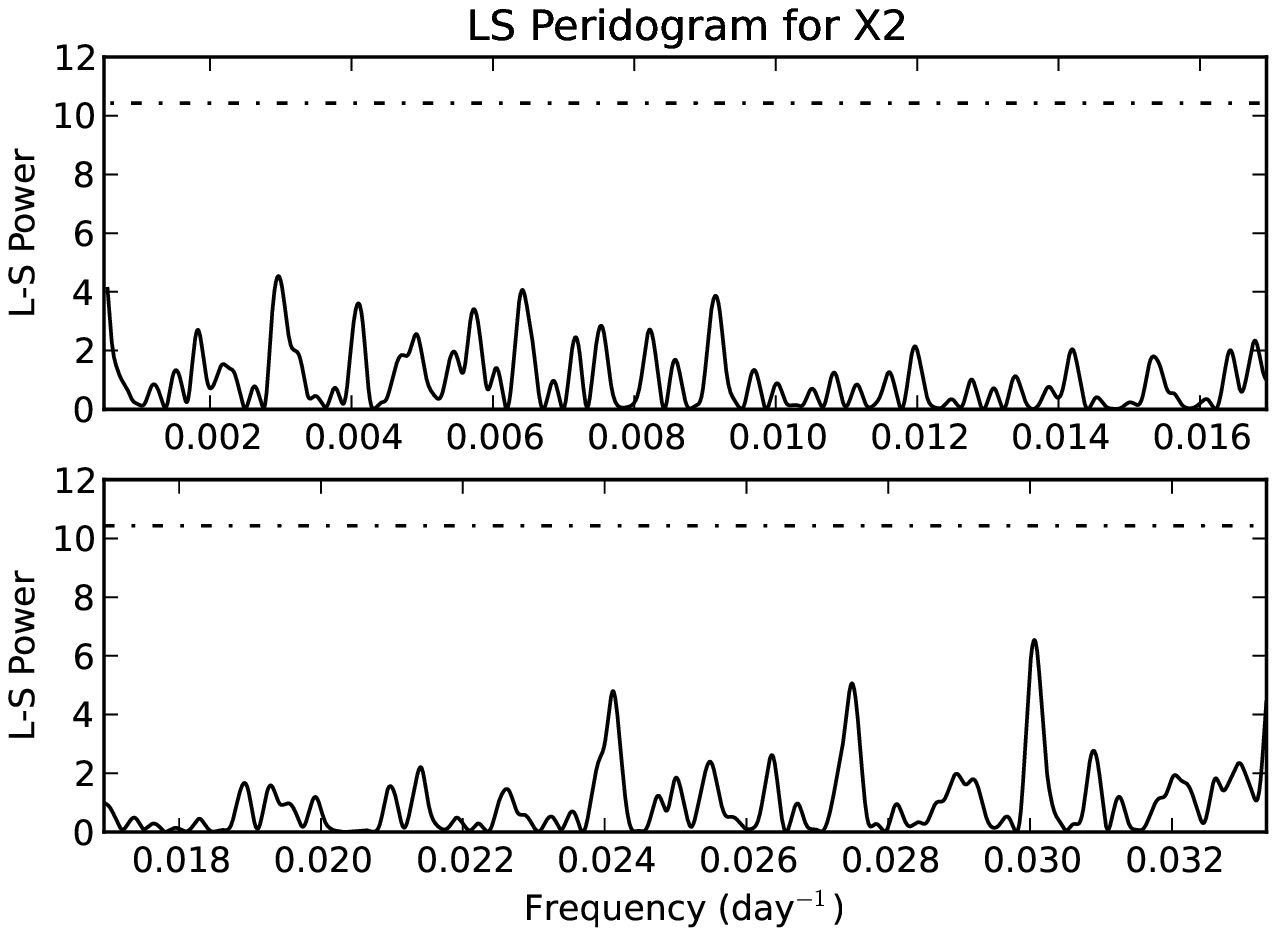}
\includegraphics[width=80mm]{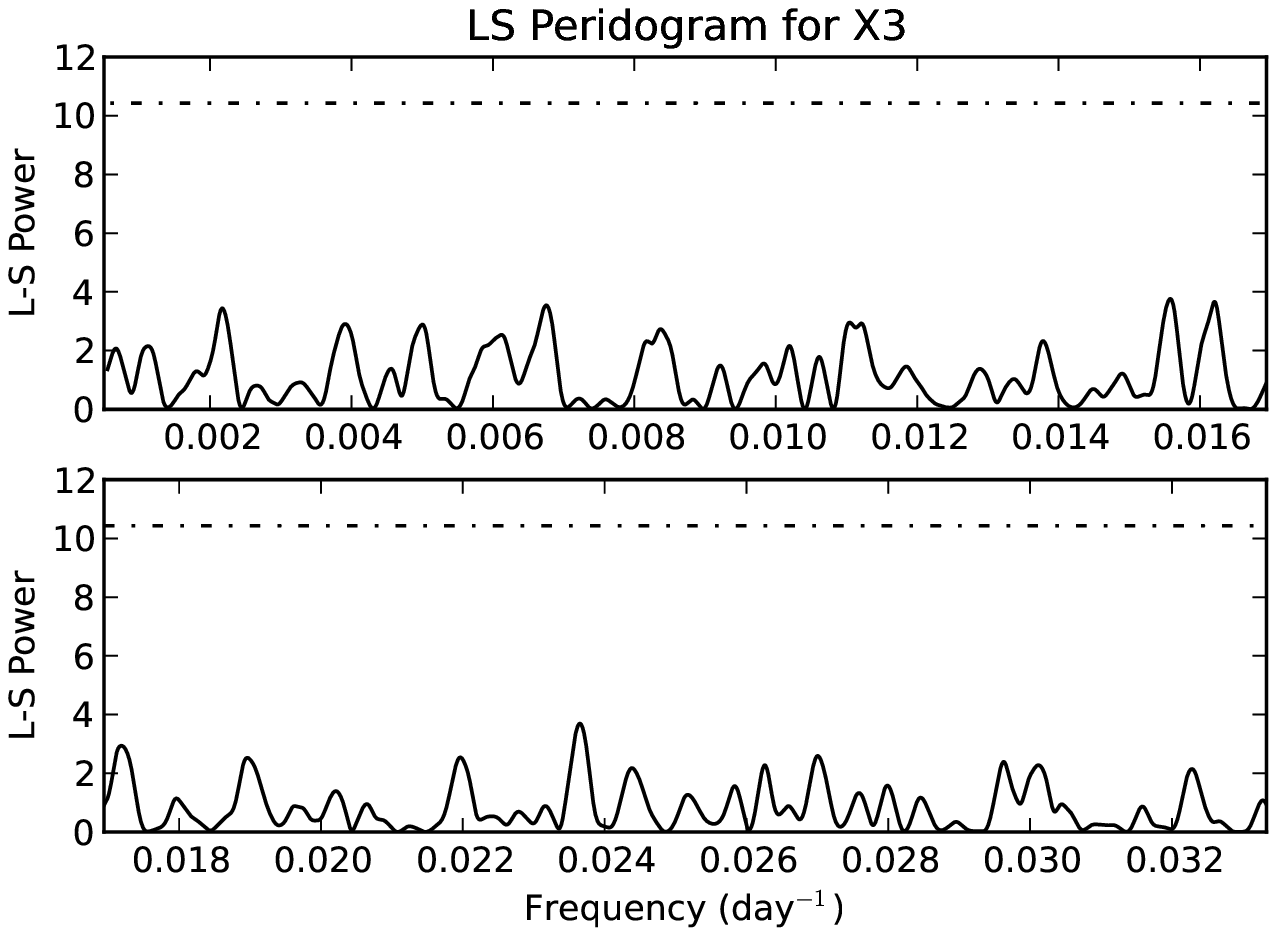}
\includegraphics[width=80mm]{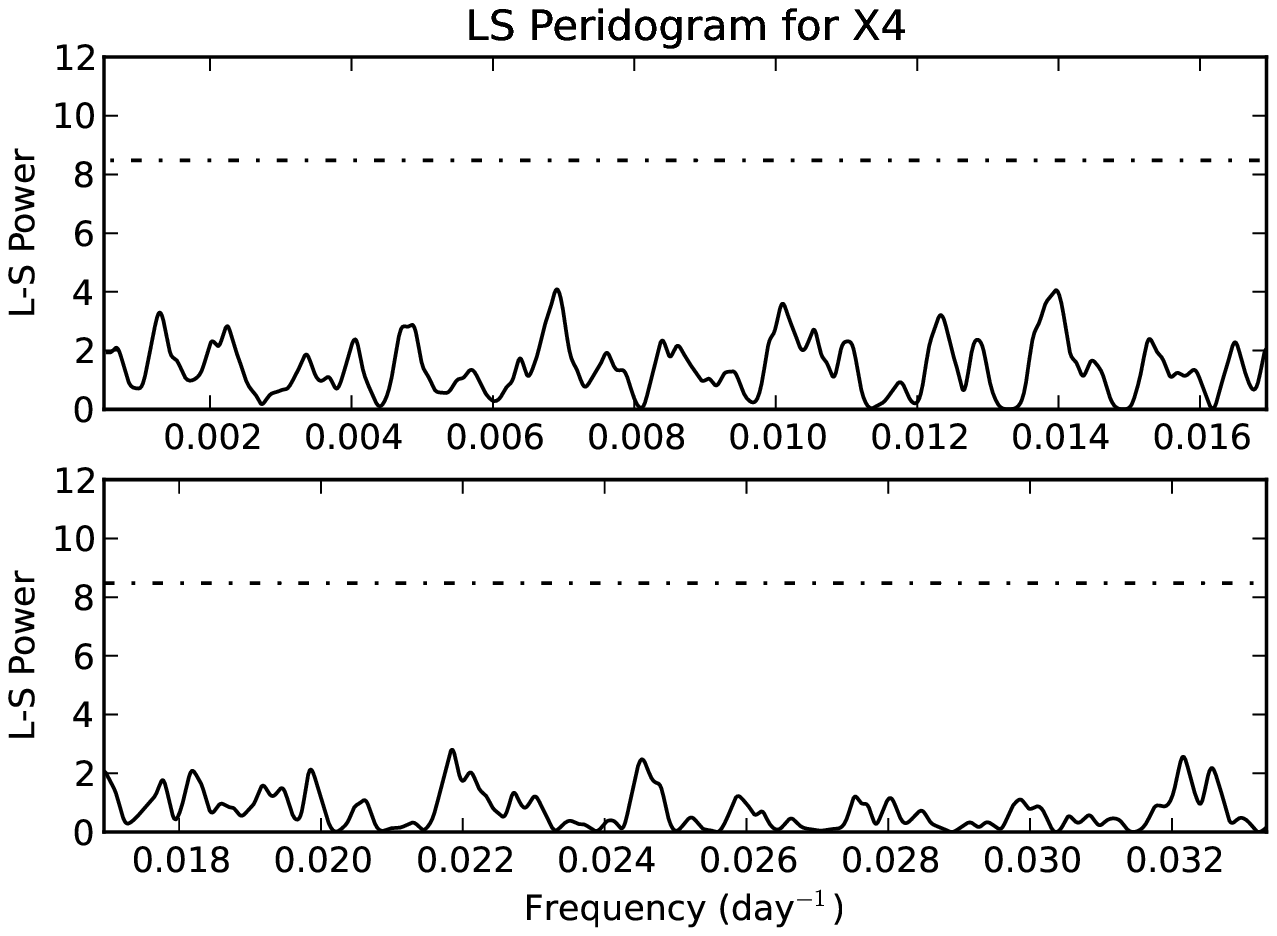}
\caption{Lomb-Scargle periodogram for the data plotted in Figure \ref{othercurves1}. The frequencies correspond to periods ranging from 30 d to 5 yr.  The dashed line represents the 3$\sigma$ Monte Carlo confidence level.}
\label{otherlsps1}
\end{figure*}

\begin{figure*}
\includegraphics[width=80mm]{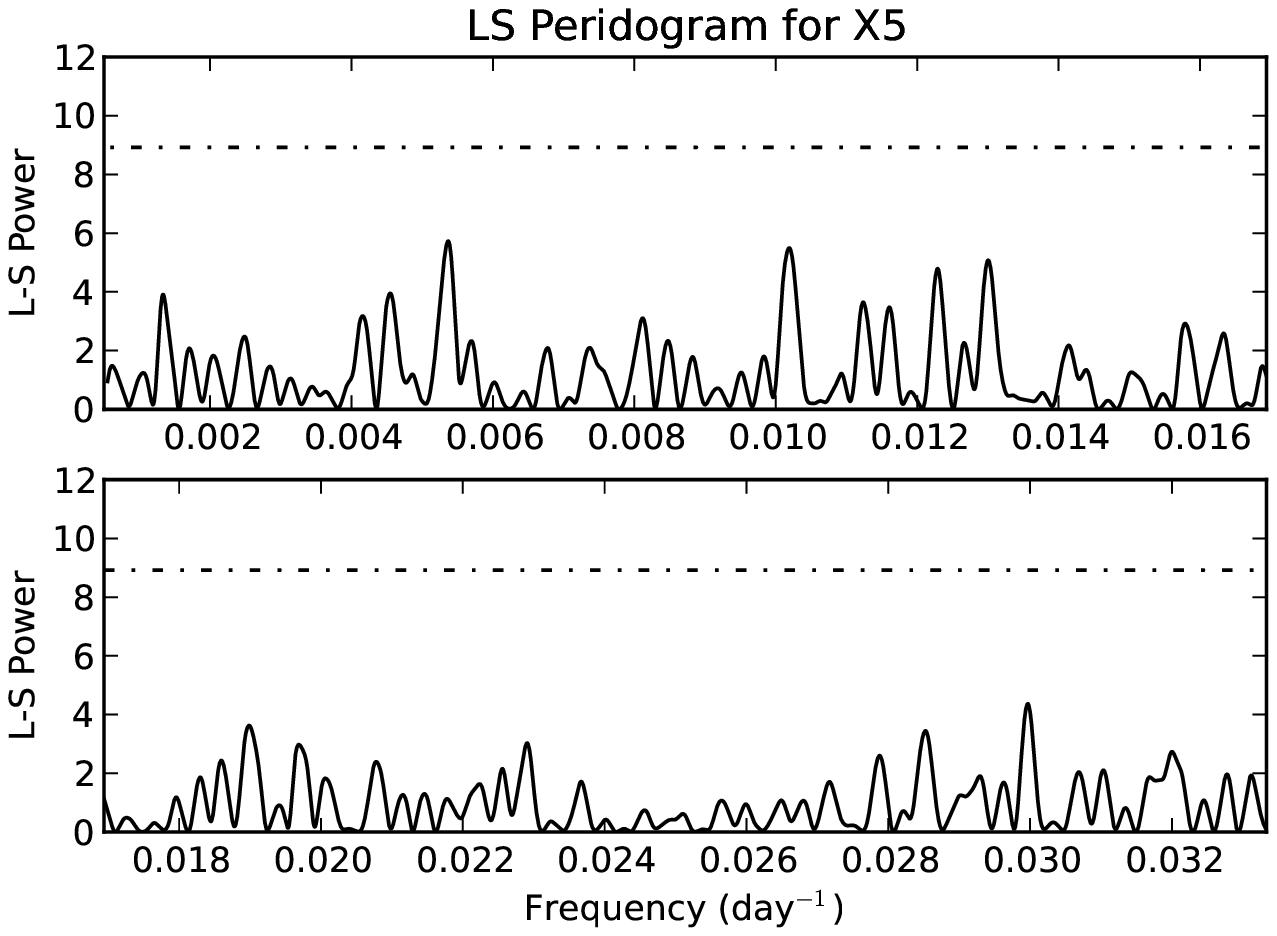}
\includegraphics[width=80mm]{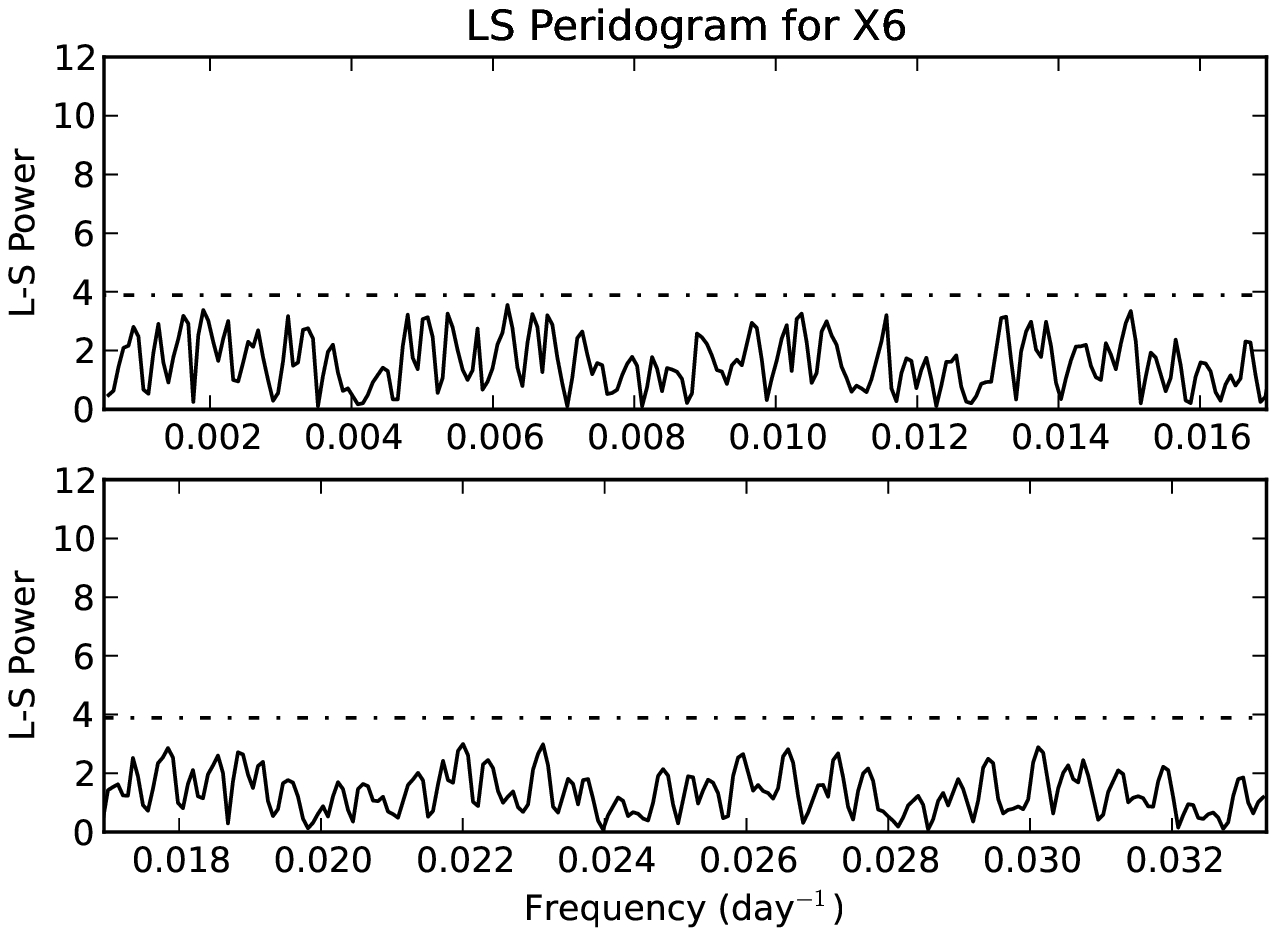}
\includegraphics[width=80mm]{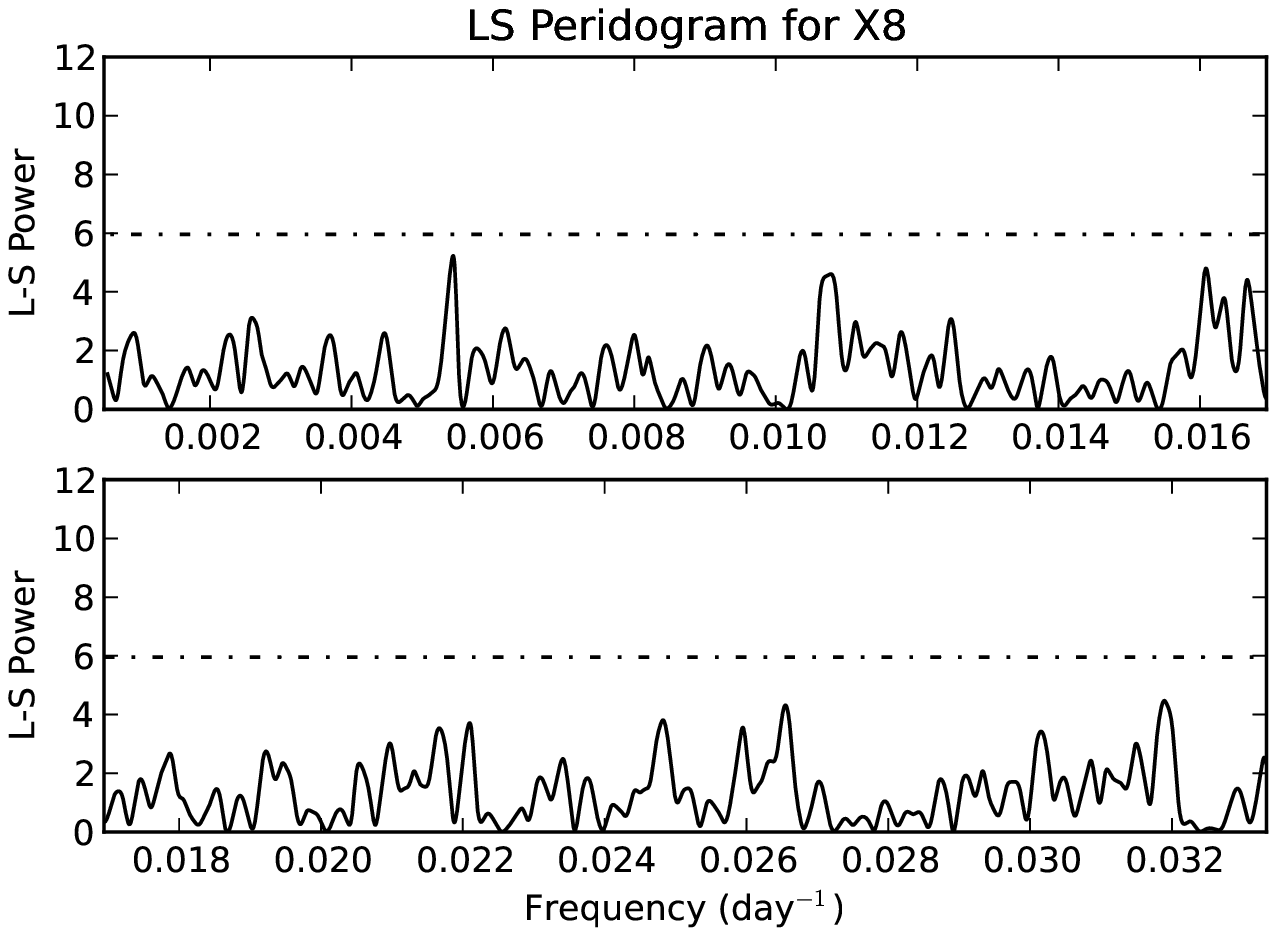}
\caption{Lomb-Scargle periodogram for the data plotted in Figure \ref{othercurves2}. The frequencies correspond to periods ranging from 30 d to 5 yr.  The dashed line represents the 3$\sigma$ Monte Carlo confidence level.}
\label{otherlsps2}
\end{figure*}

\section{Window Functions}

\begin{figure*}
\includegraphics[width=80mm]{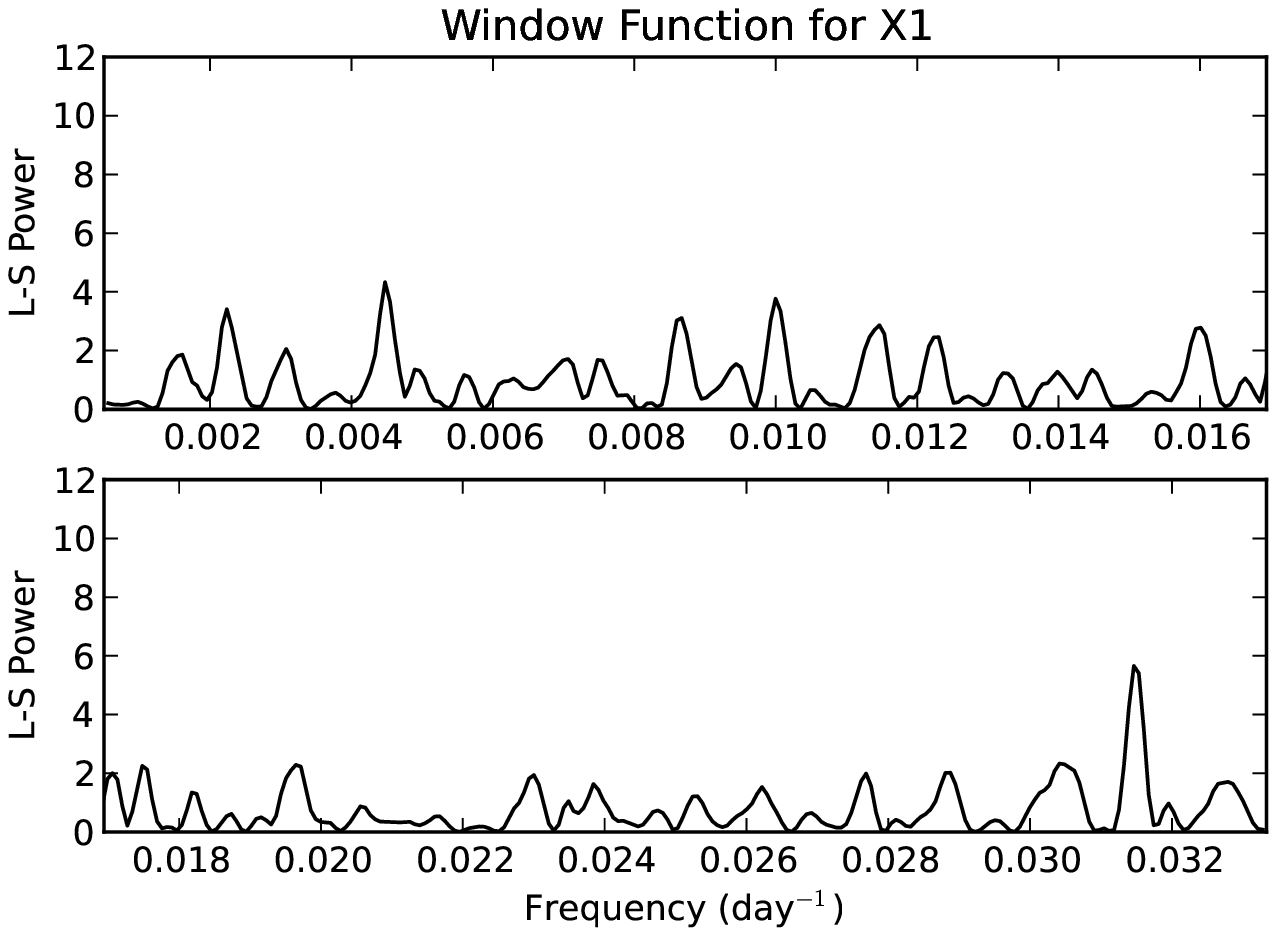}
\includegraphics[width=80mm]{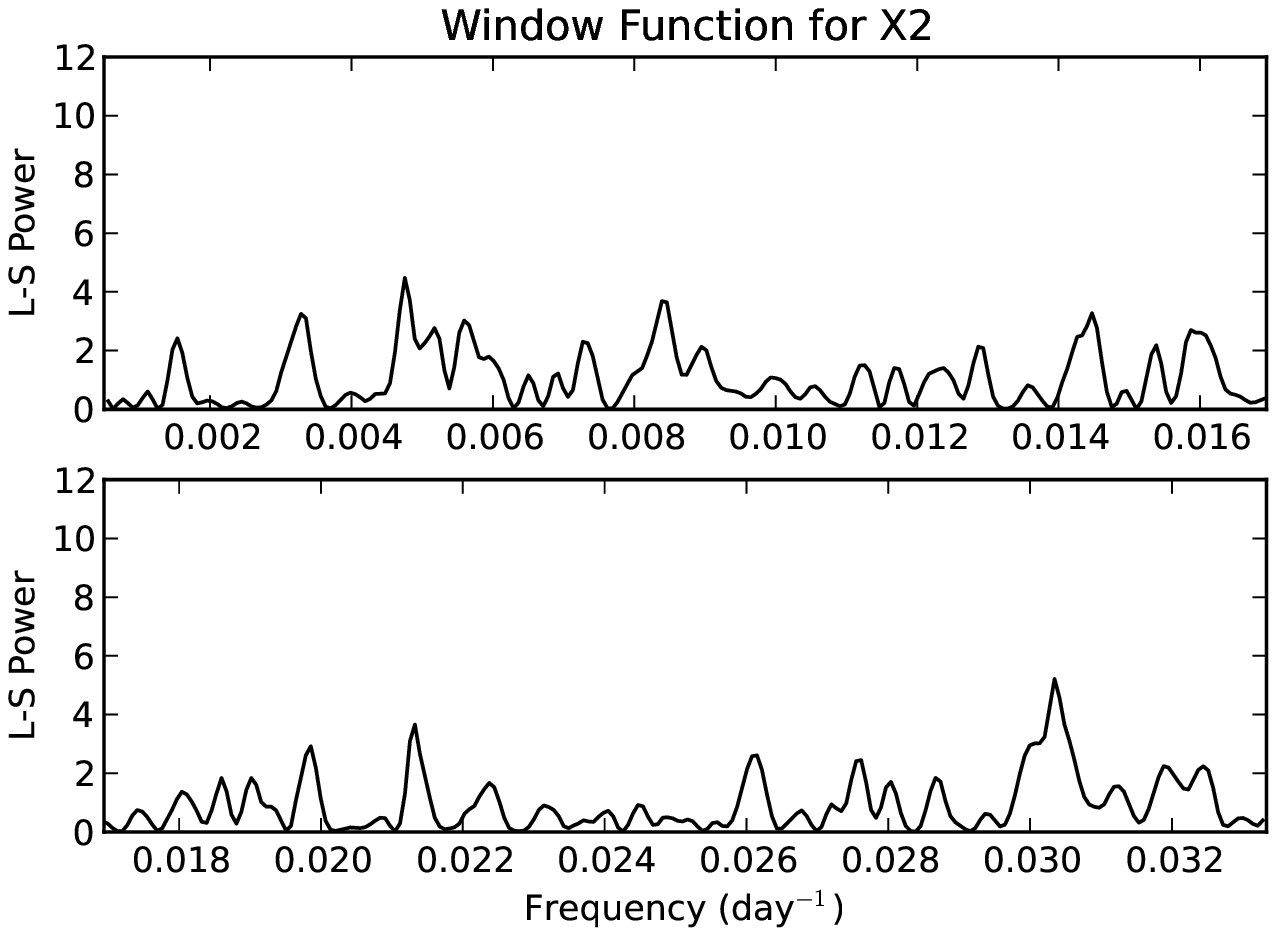}
\includegraphics[width=80mm]{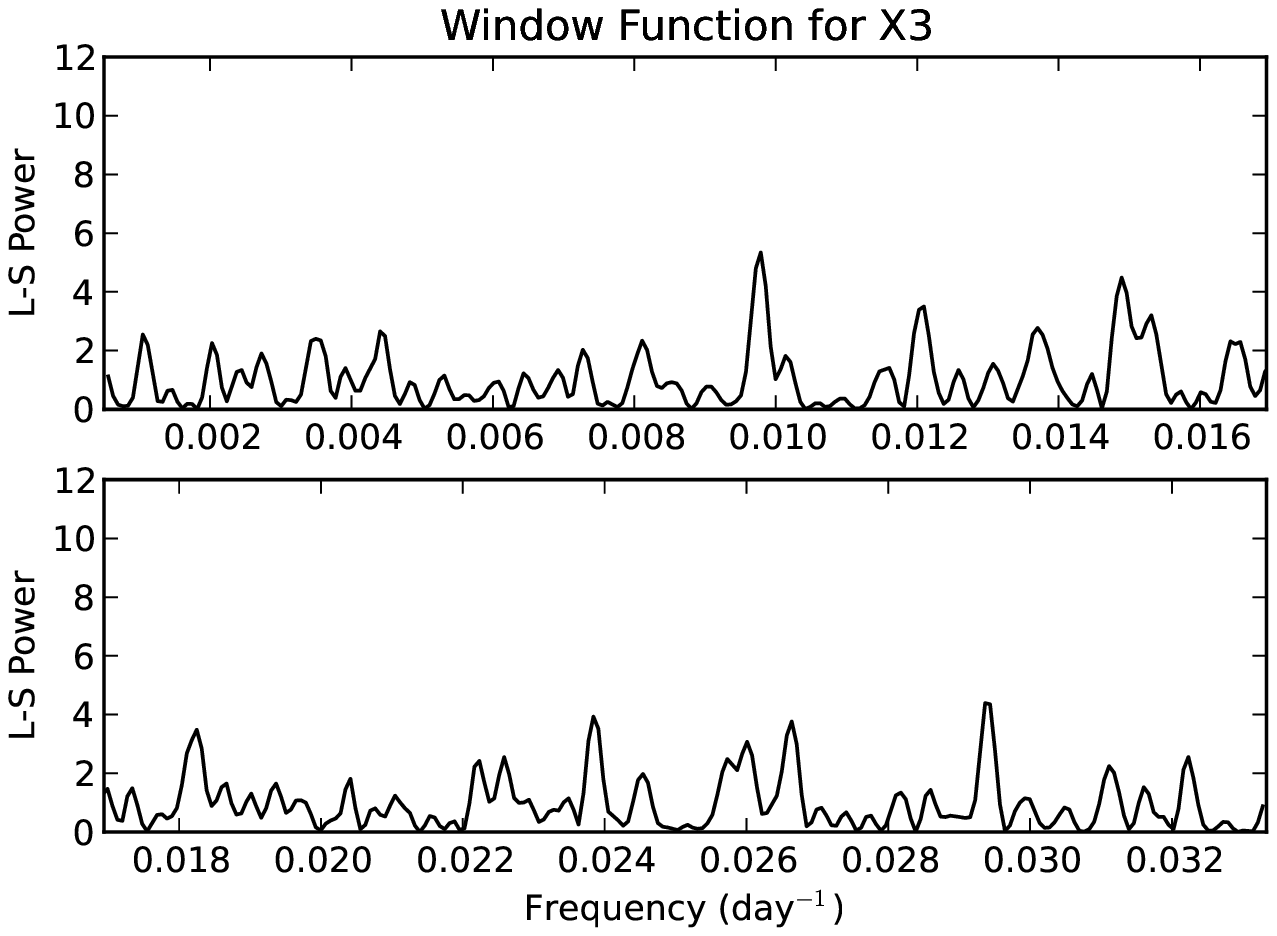}
\includegraphics[width=80mm]{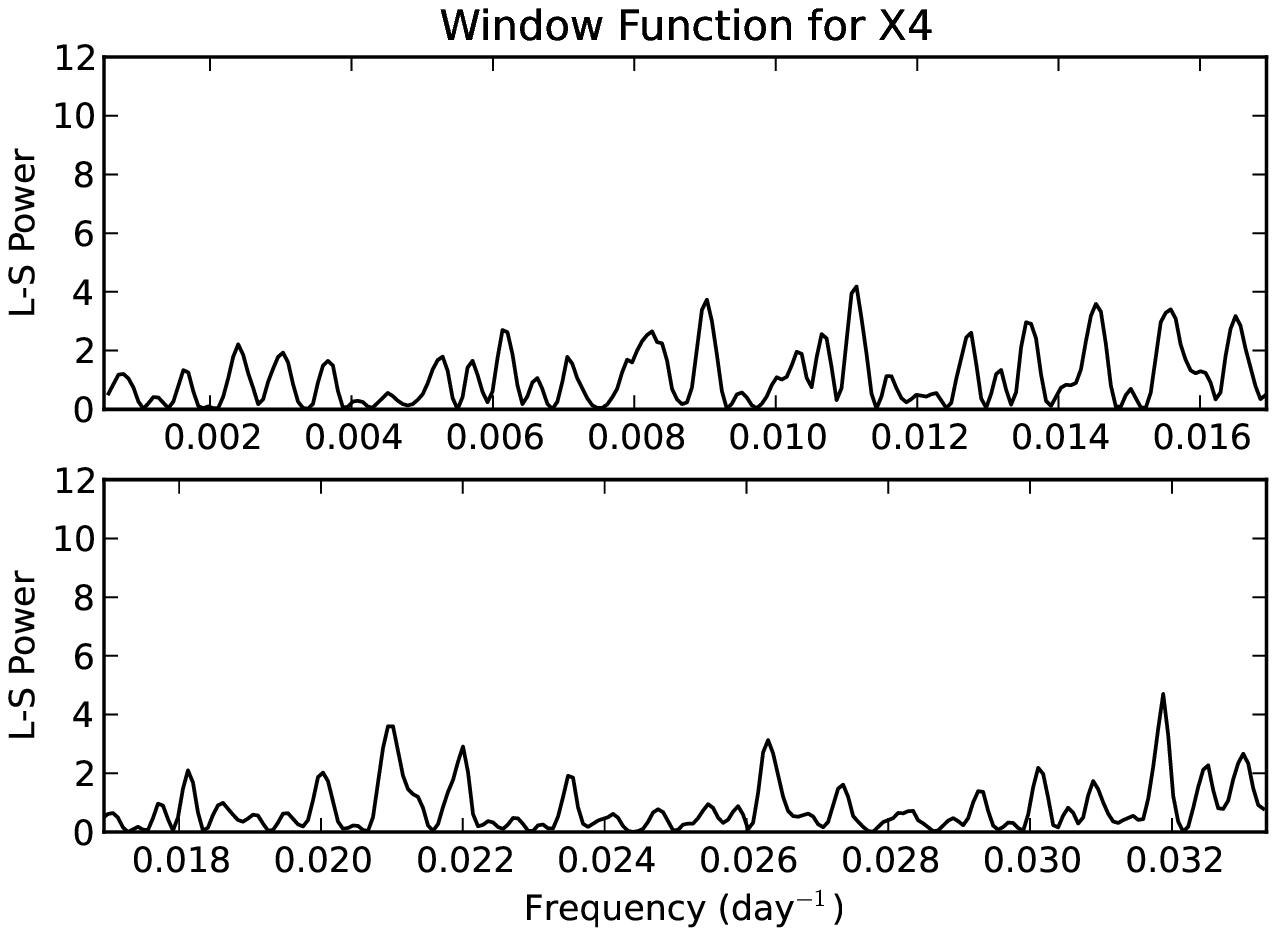}
\caption{Window functions for the data plotted in Figure \ref{othercurves1}. The frequencies correspond to periods ranging from 30 d to 5 yr.}
\label{windows1}
\end{figure*}

\begin{figure*}
\includegraphics[width=80mm]{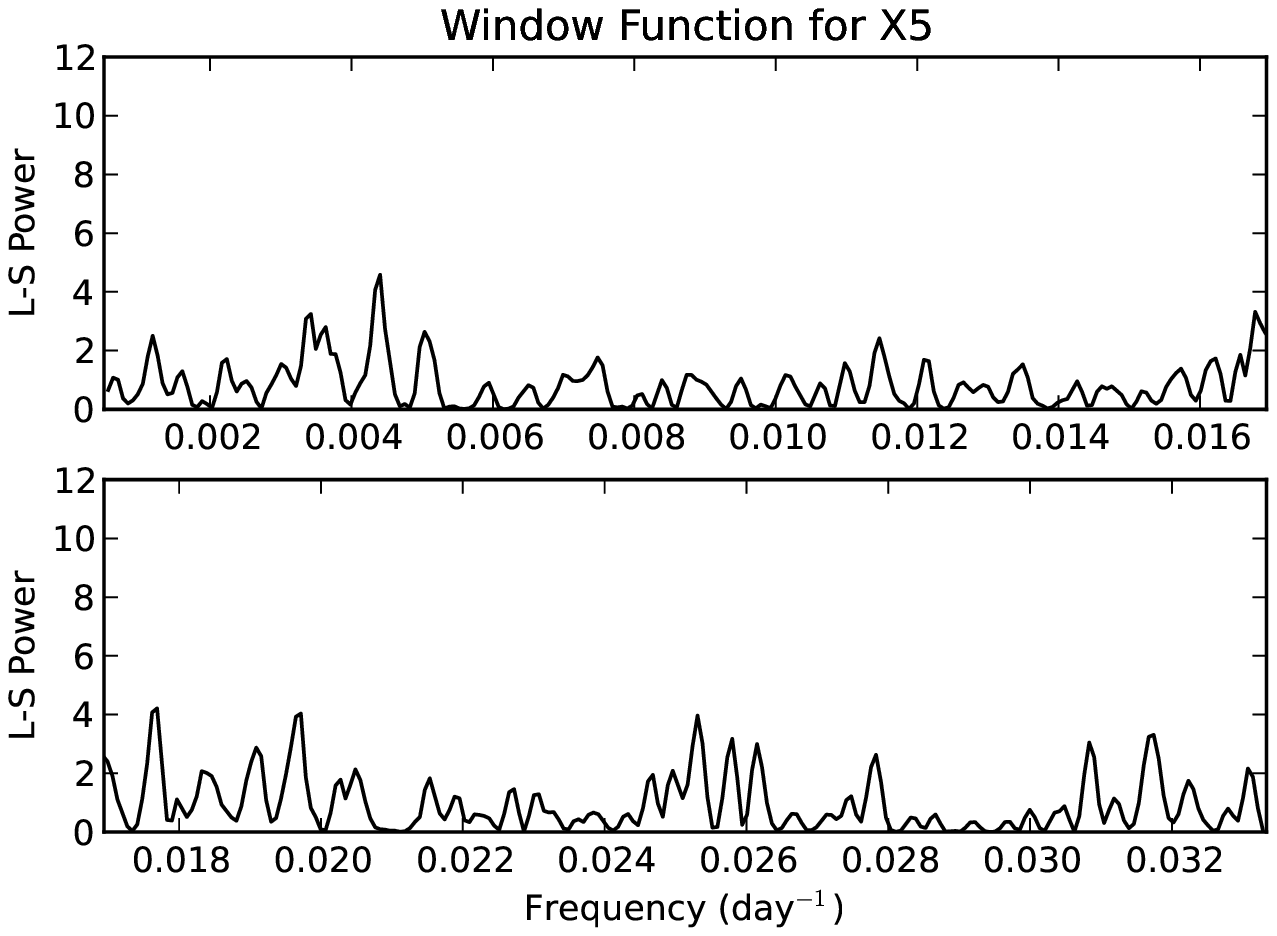}
\includegraphics[width=80mm]{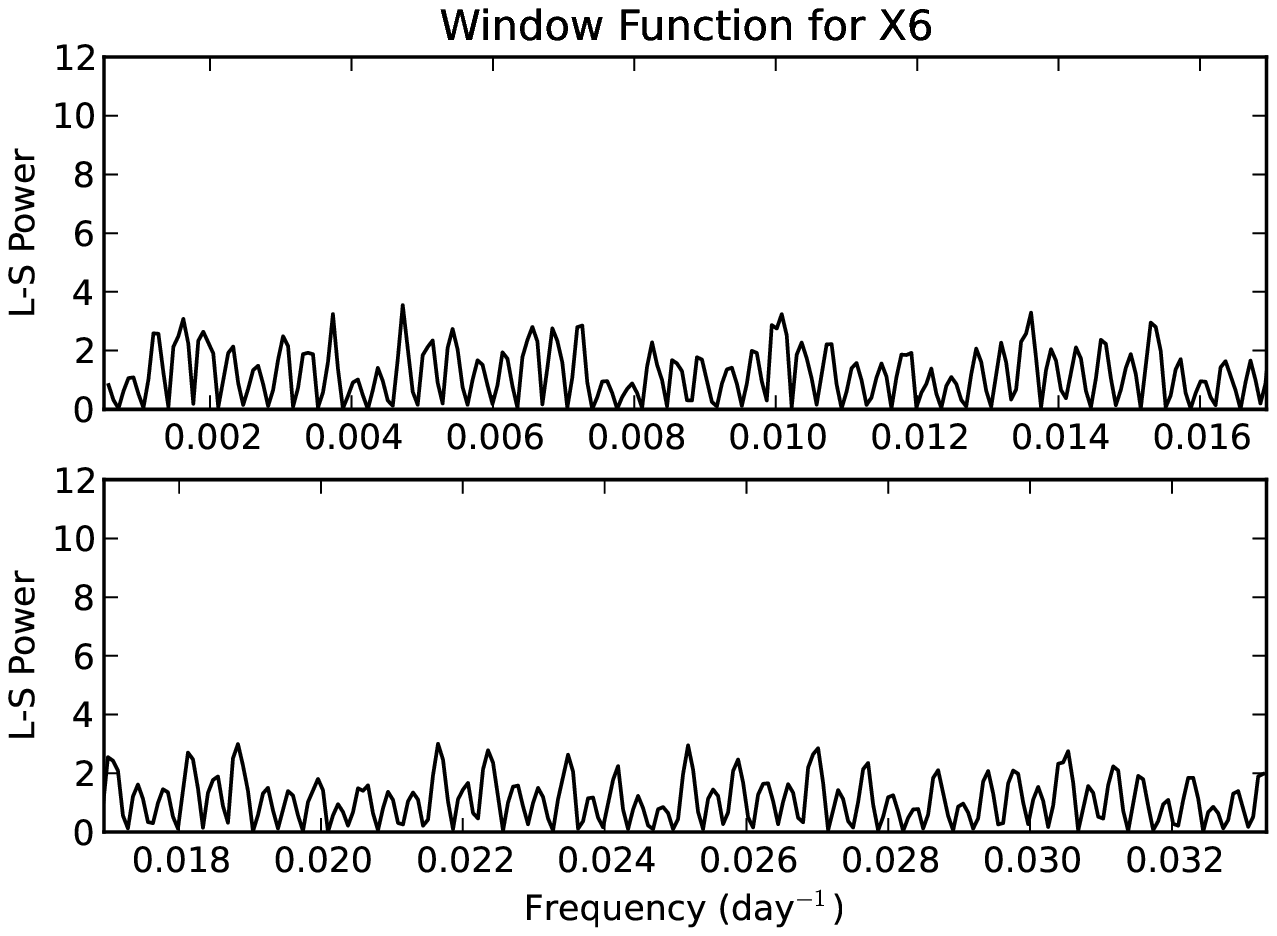}
\includegraphics[width=80mm]{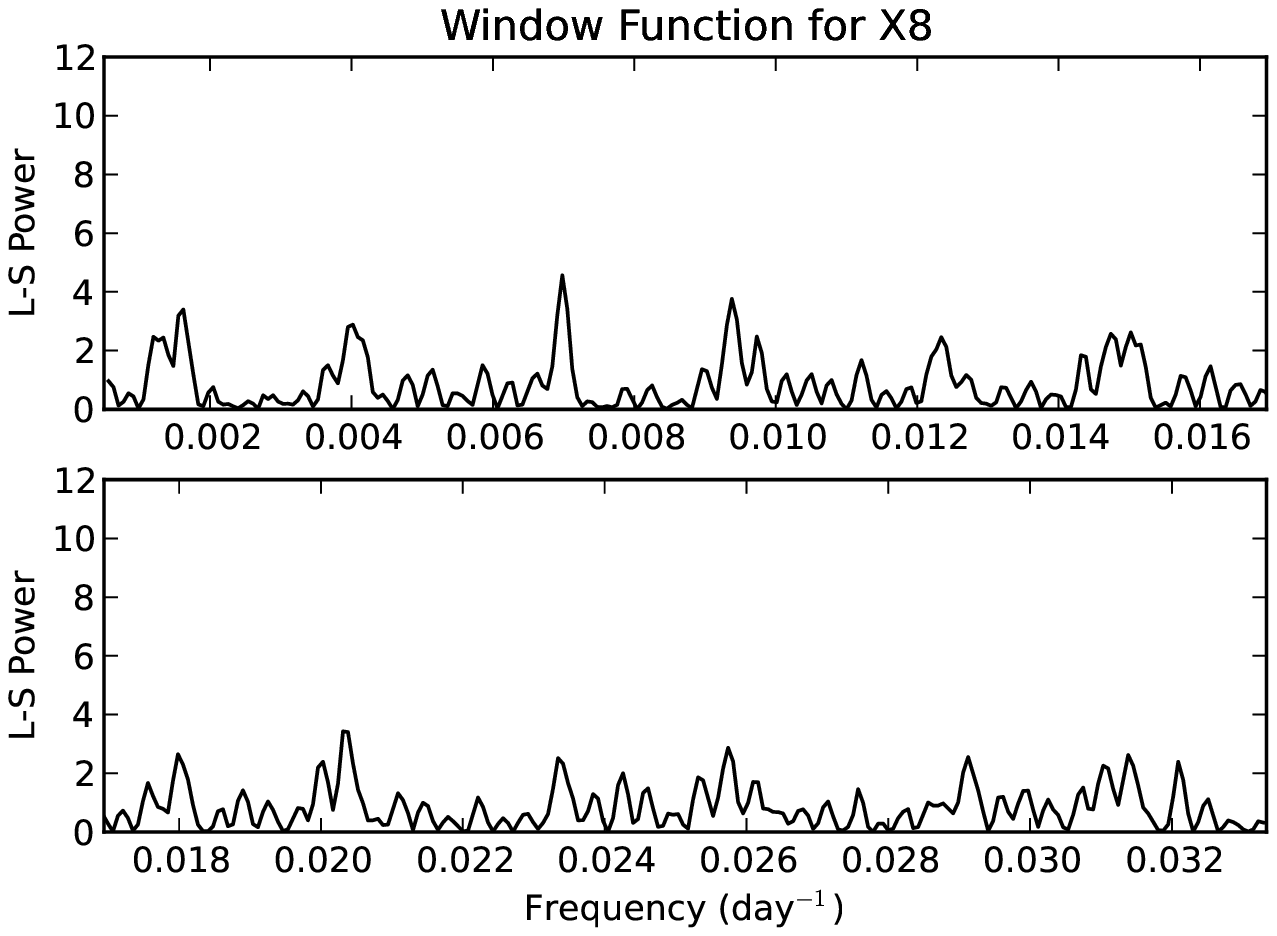}
\caption{Window functions for the data plotted in Figure \ref{othercurves2}. The frequencies correspond to periods ranging from 30 d to 5 yr.}
\label{windows2}
\end{figure*}

\section{Period Detection Sensitivity}

\begin{figure*}
\includegraphics[width=130mm, angle=90]{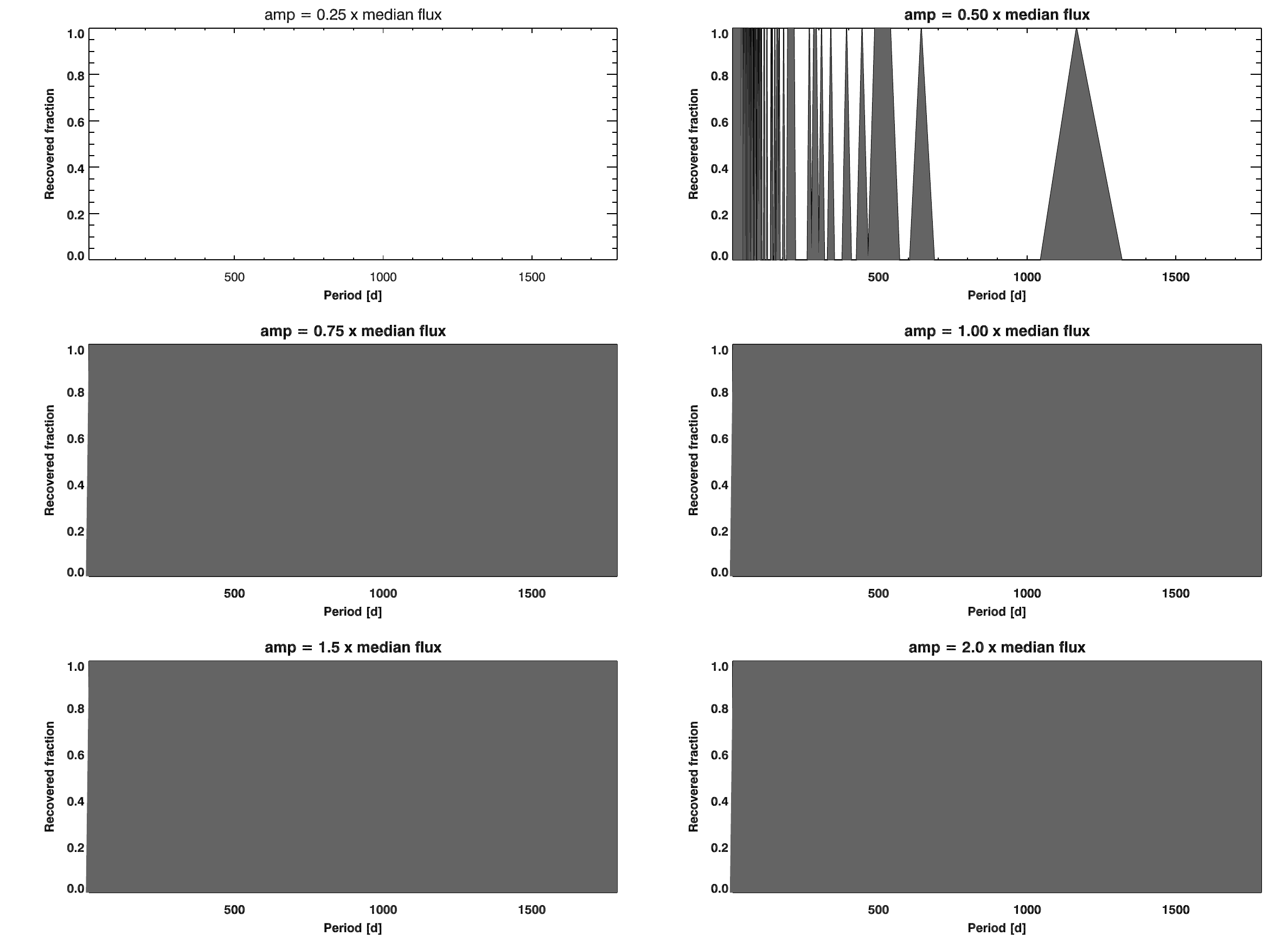}
\caption{Period detection sensitivity for source X1. We use the times and error bars of the extracted lightcurve to generate 10,000 simulated lightcurves per period bin with randomized phases. LSP analysis was performed to test whether the input periods are recovered.  The recovery fraction of each input period is determined by evaluating the LSPs of all the simulated lightcurves in each period bin. Each panel represents a different fixed input modulation amplitude.  The recovery fraction is indicated by grey shading.  Agreement to within 10 per cent of the input period is deemed a successful recovery.}
\label{recover1}
\end{figure*}

\begin{figure*}
\includegraphics[width=130mm, angle=90]{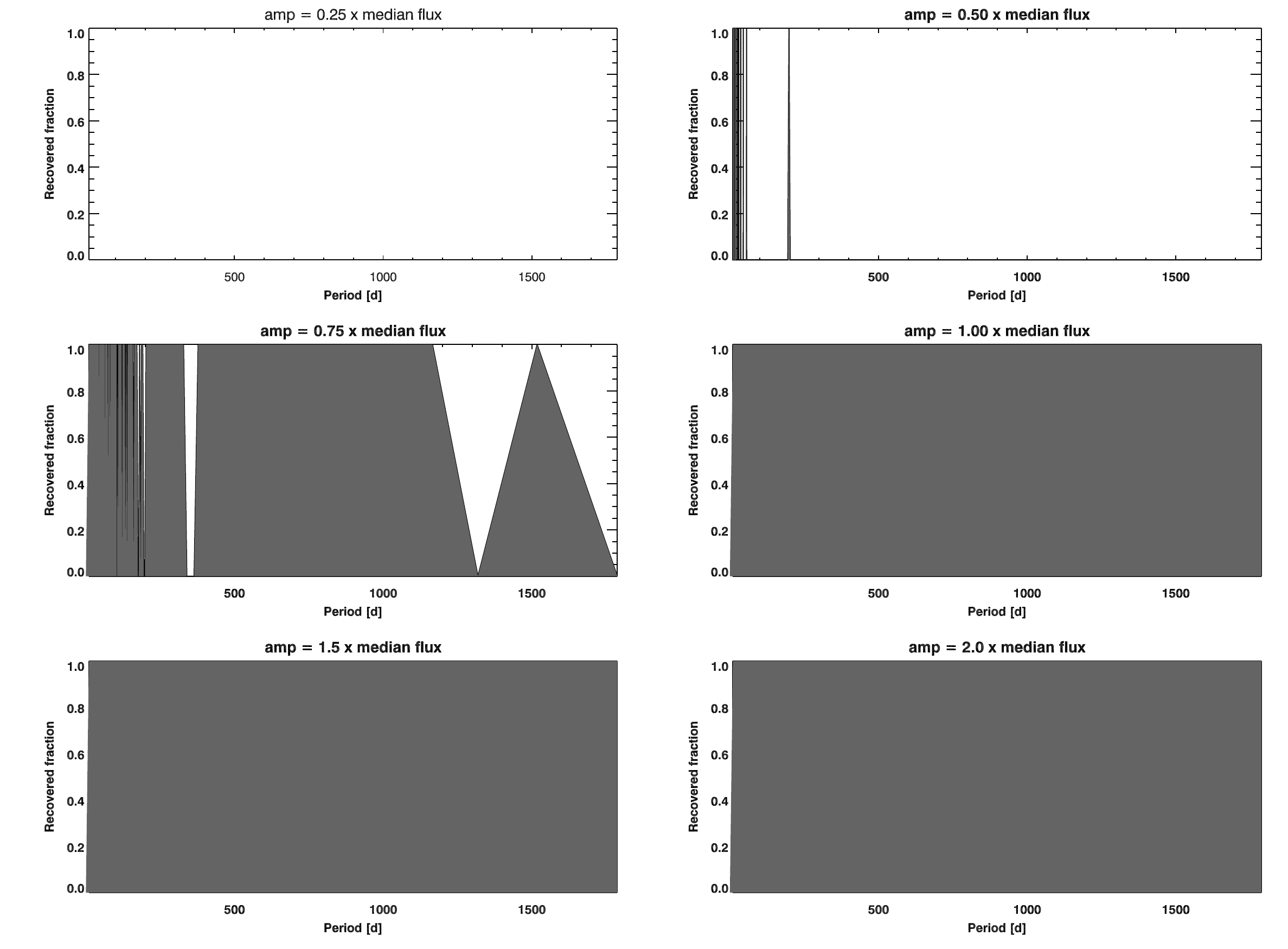}
\caption{Period detection sensitivity for source X2. We use the times and error bars of the extracted lightcurve to generate 10,000 simulated lightcurves per period bin with randomized phases. LSP analysis was performed to test whether the input periods are recovered.  The recovery fraction of each input period is determined by evaluating the LSPs of all the simulated lightcurves in each period bin. Each panel represents a different fixed input modulation amplitude.  The recovery fraction is indicated by grey shading.  Agreement to within 10 per cent of the input period is deemed a successful recovery.}
\label{recover2}
\end{figure*}

\begin{figure*}
\includegraphics[width=130mm, angle=90]{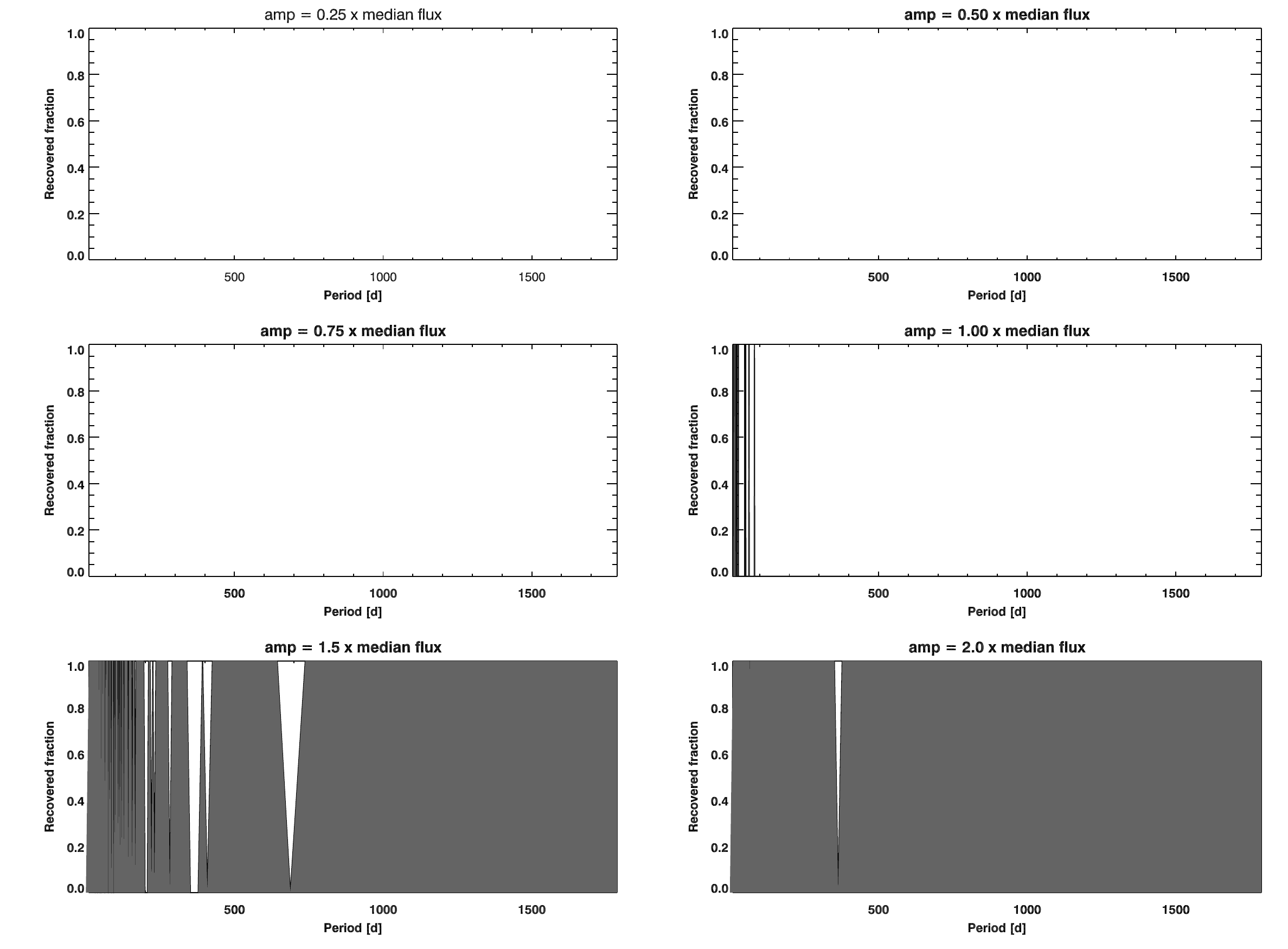}
\caption{Period detection sensitivity for source X3. We use the times and error bars of the extracted lightcurve to generate 10,000 simulated lightcurves per period bin with randomized phases. LSP analysis was performed to test whether the input periods are recovered.  The recovery fraction of each input period is determined by evaluating the LSPs of all the simulated lightcurves in each period bin. Each panel represents a different fixed input modulation amplitude.  The recovery fraction is indicated by grey shading.  Agreement to within 10 per cent of the input period is deemed a successful recovery.}
\label{recover3}
\end{figure*}

\begin{figure*}
\includegraphics[width=130mm, angle=90]{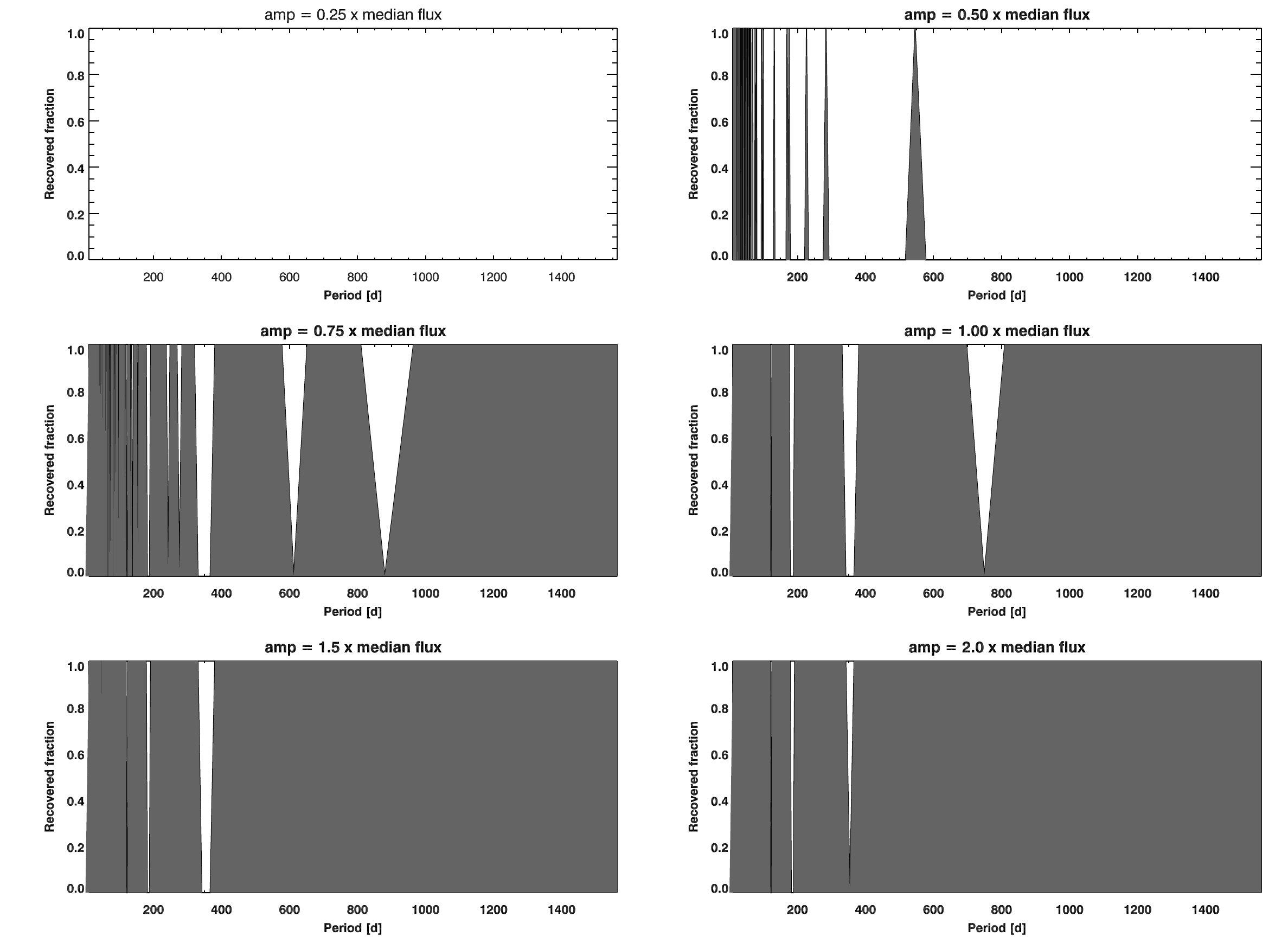}
\caption{Period detection sensitivity for source X4. We use the times and error bars of the extracted lightcurve to generate 10,000 simulated lightcurves per period bin with randomized phases. LSP analysis was performed to test whether the input periods are recovered.  The recovery fraction of each input period is determined by evaluating the LSPs of all the simulated lightcurves in each period bin. Each panel represents a different fixed input modulation amplitude.  The recovery fraction is indicated by grey shading.  Agreement to within 10 per cent of the input period is deemed a successful recovery.}
\label{recover5}
\end{figure*}

\begin{figure*}
\includegraphics[width=130mm, angle=90]{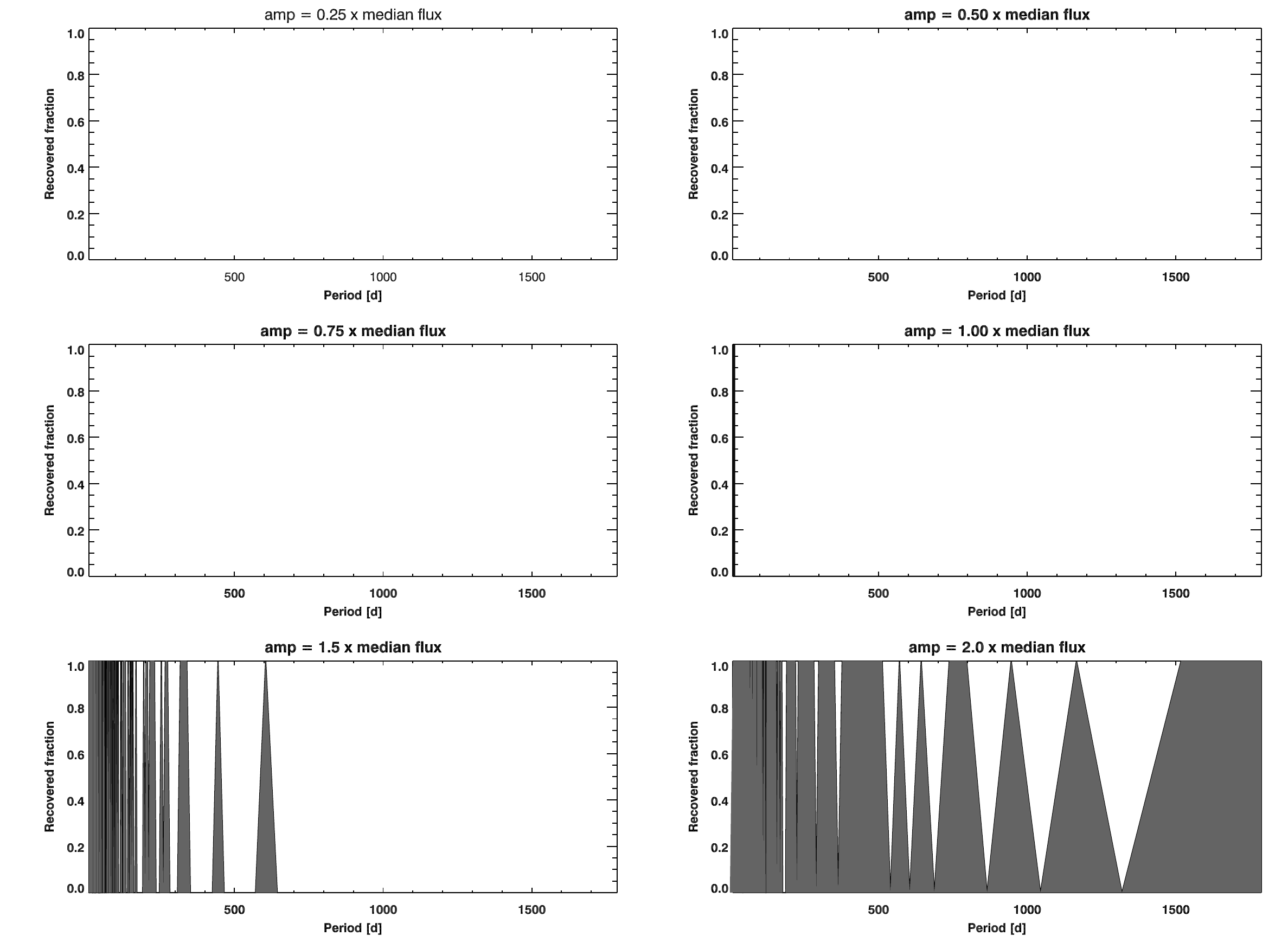}
\caption{Period detection sensitivity for source X5. We use the times and error bars of the extracted lightcurve to generate 10,000 simulated lightcurves per period bin with randomized phases. LSP analysis was performed to test whether the input periods are recovered.  The recovery fraction of each input period is determined by evaluating the LSPs of all the simulated lightcurves in each period bin. Each panel represents a different fixed input modulation amplitude.  The recovery fraction is indicated by grey shading.  Agreement to within 10 per cent of the input period is deemed a successful recovery.}
\label{recover6}
\end{figure*}

\begin{figure*}
\includegraphics[width=130mm, angle=90]{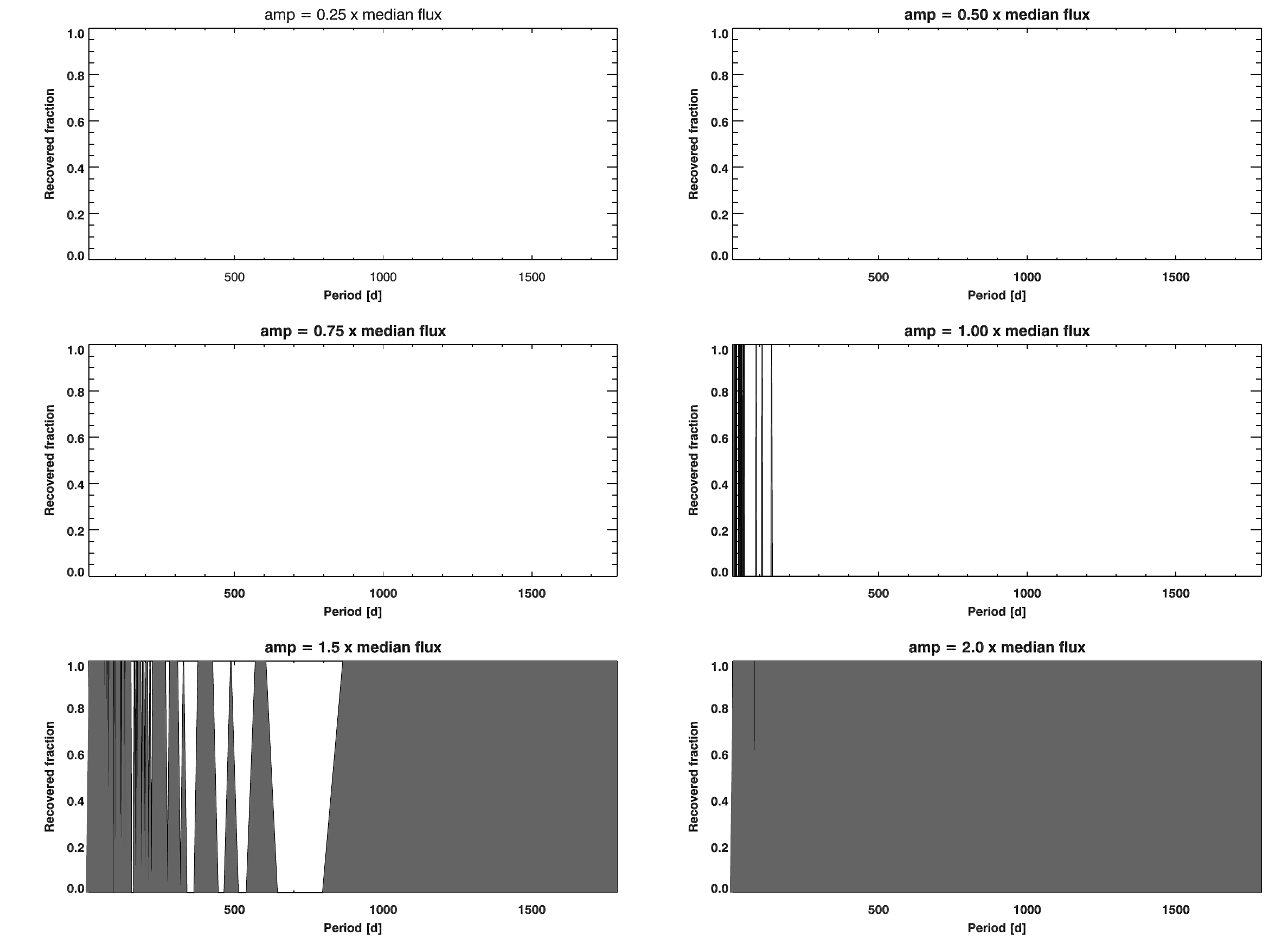}
\caption{Period detection sensitivity for source X6. We use the times and error bars of the extracted lightcurve to generate 10,000 simulated lightcurves per period bin with randomized phases. LSP analysis was performed to test whether the input periods are recovered.  The recovery fraction of each input period is determined by evaluating the LSPs of all the simulated lightcurves in each period bin. Each panel represents a different fixed input modulation amplitude.  The recovery fraction is indicated by grey shading.  Agreement to within 10 per cent of the input period is deemed a successful recovery.}
\label{recover7}
\end{figure*}

\begin{figure*}
\includegraphics[width=130mm, angle=90]{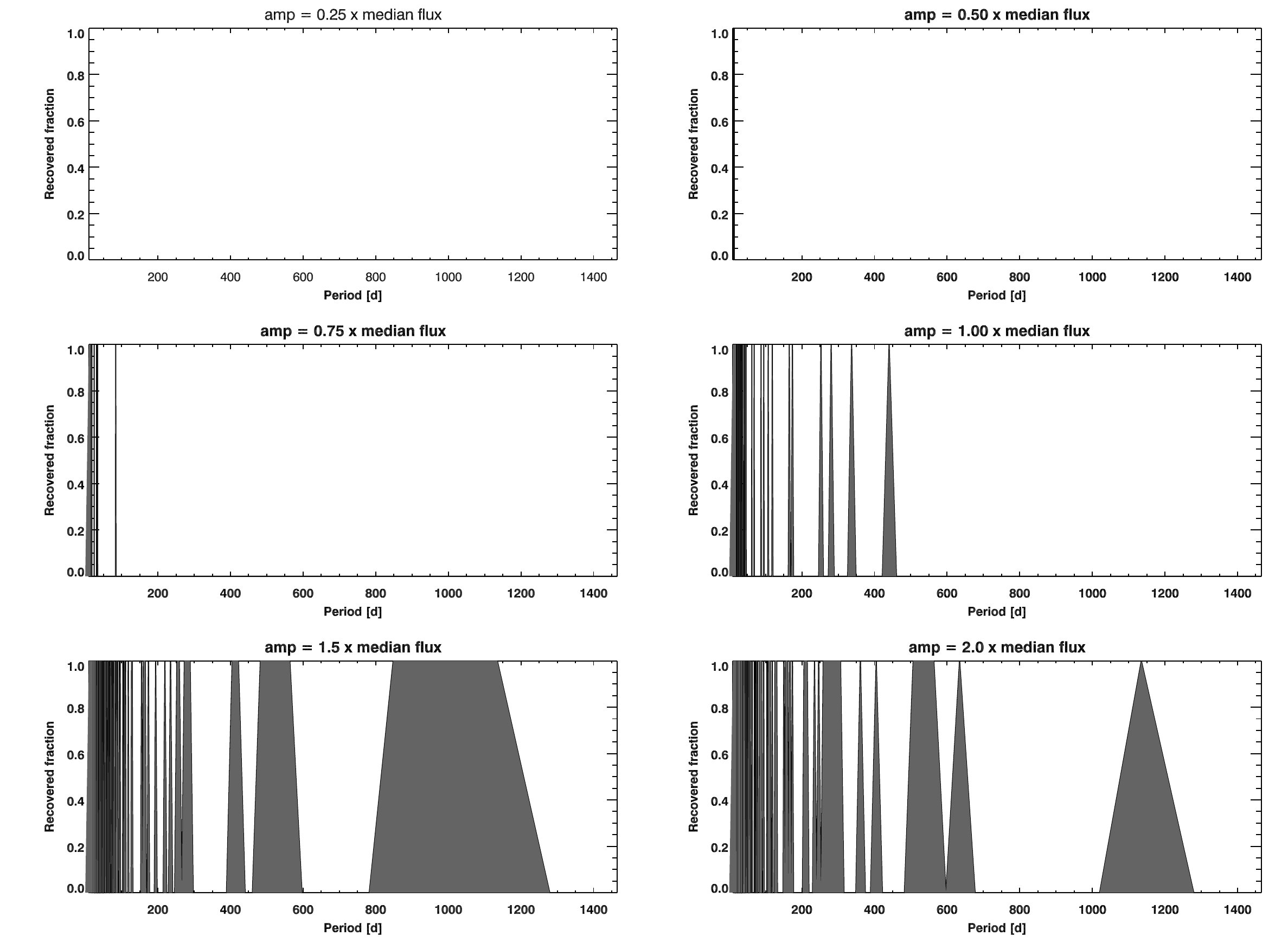}
\caption{Period detection sensitivity for source X8. We use the times and error bars of the extracted lightcurve to generate 10,000 simulated lightcurves per period bin with randomized phases. LSP analysis was performed to test whether the input periods are recovered.  The recovery fraction of each input period is determined by evaluating the LSPs of all the simulated lightcurves in each period bin. Each panel represents a different fixed input modulation amplitude.  The recovery fraction is indicated by grey shading.  Agreement to within 10 per cent of the input period is deemed a successful recovery.}
\label{recover4}
\end{figure*}

\bsp

\label{lastpage}

\end{document}